\documentclass[12pt]{article}

\usepackage[dvips]{graphicx}
\usepackage{amssymb,amsthm,amsmath,latexsym}

\newcommand{\od}{\overset{\mbox{\rm def}}=}

\theoremstyle{definition}
\newtheorem{definition}{Definition}

\theoremstyle{plain}
\newtheorem{lemma}{Lemma}

\theoremstyle{plain}
\newtheorem{theorem}{Theorem}

\theoremstyle{plain}
\newtheorem{corollary}{Corollary}

\theoremstyle{definition}
\newtheorem{remark}{Remark}

\theoremstyle{definition}
\newtheorem{example}{Example}

\begin{document}

\title{Weak asymptotic solution of the phase field system
in the case of confluence of free boundaries
in the Stefan problem with undercooling}
\author{V.G.~Danilov\thanks{Moscow Technical University of
Communication and Informatics,\hfill\break 
e-mail: danilov@miem.edu.ru
\hfill\break
This work was supported by the Russian Foundation for Basic
Research under grant 05-01-00912.}}
\date{}
\maketitle

\abstract{We assume that
the Stefan problem with undercooling
has a classical solution
until the moment of contact of free boundaries
and the free boundaries have continuous velocities 
until the moment of contact.
Under these assumptions, we construct a smooth
approximation of the global solution of the Stefan problem
with undercooling, which, until the contact, gives
the classical solution mentioned above and,
after the contact, gives a solution which is the solution of
the heat equation.}

\section{Introduction}\label{S:intro}

In this paper, we study the confluence of free boundaries
in the Stefan problem with undercooling
in the one-dimensional case.
We at once note that
an analysis of the one-dimensional problem
is not the final goal but is only a necessary step
in the study of the multidimensional problem.

In the preset paper, we study the problem
in the domain
$Q=\Omega\times[0,t_{1}]$,
where $\Omega=[l_{1},l_{2}]$ is an interval.
We assume that the interval $\Omega$
is divided into three parts $\Omega^{+}_{l,r}(t)$
and $\Omega^{-}(t)$ as follows:
\begin{equation*}
\Omega^{+}_{l}(t)=[l_{1},\varphi_{1}(t)], \quad
\Omega^{-}(t)=[\varphi_{1}(t),\varphi_{2}(t)], \quad
\Omega^{+}_{r}(t)=[\varphi_{2}(t), l_{2}],
\end{equation*}
where $\varphi_{i}(t)$, $i=1,2$,
are the free boundaries of phases "$+$" and "$-$".
We assume that the phase~"$+$" occupies the intervals
$\Omega^{+}_{l,r}(t)$
and the phase~"$-$" occupies the interval $\Omega^{-}(t)$.

We shall construct a smooth approximation of solutions
of the Stefan problem with undercooling
under the assumption that the motion of the free boundary
is the motion of the front of a nonlinear wave
and the confluence of free boundaries
is interaction of solitary nonlinear waves.
The possibility of this interpretation is given  by
the models of phase field~\cite{12} proposed by G.~Caginalp.
In fact, the choice of the method for approximating
the limit Stefan problems with undercooling is unessential for
us, because we do not prove that the approximations thus
constructed are close to the corresponding solutions
of the phase field system.

For example, we could use the definition of the generalized
solution of the limit problems including the order function
(the nonlinear wave)~\cite{111}.

Here our considerations are based on the following simple fact.
Suppose that there are two families of solutions
(exact and approximate solutions)
of some problem depending on a small parameter~$\varepsilon$.
Suppose that both these families have the properties that
permit passing to the limit as $\varepsilon\to 0$
in the weak sense.
Suppose also that the family of approximate solutions
satisfies a problem with a right-hand side
small as $\varepsilon\to 0$ in the weak sense.
Then the weak limits of both families are solutions
of the same limit problem and
if the latter has a unique solution, then the difference
between the exact and approximate solutions tends to~$0$
as $\varepsilon\to0$ at least in the weak sense.

We recall that the phase field system has the form
\begin{equation}\label{PF:1}
L\theta= -\frac{\partial u}{\partial t},\qquad
\varepsilon Lu-\frac{u-u^{3}}{\varepsilon}-\theta=0,
\end{equation}
where
\begin{equation*}\label{L:1}
L=\frac{\partial}{\partial t}-\frac{\partial^{2}}{\partial x^{2}},
\qquad x\in [l_{1},l_{2}], \quad t\in[0,t_{1}].
\end{equation*}

The function $\theta=\theta(x,t,\varepsilon)$ has the meaning of
the temperature,
and the function $u=u(x,t,\varepsilon)$,
which is called the order function,
determined the phase state of the medium:
$u\simeq-1$ corresponds to the phase~"$-$"
in the domain $\Omega^{-}(t)$,
and $u\simeq 1$ corresponds to the phase~"$+$"
in the domains  $\Omega^{-}_{l,r}(t)$.

Passing to the limit as $\varepsilon\to 0$ in (\ref{PF:1}),
we obtain the Stefan problem with undercooling.

This passage to the limit is possible, for example, in the case
where the corresponding limit problems have classical solutions.
In this case, the weak limits as $\varepsilon\to 0$ of solutions
(\ref{PF:1}) give these solutions~\cite{12,2,3}.

The smooth approximations of solutions of the Stefan problem
with kinetic undercooling constructed in this paper
are approximate (in the above sense) solutions
of the phase field system
and they admit a weak passage to the limit as
$\varepsilon\to 0$.
In this case, we obtain the limit problems (and their solutions)
describing the process of confluence of free boundaries.
To construct these approximations, we use the assumption
that the classical sharp fronted solution of the Stefan problem
with kinetic undercooling exists
until the confluence of free boundaries begins.
This is a natural assumption in our paper
(otherwise,
it is not clear the confluence of what is considered)
and can be proved\footnote{A.~M.~Meirmanov.
A private communication.}.

We repeat that, under this assumption, we construct
an approximation of solution of the limit problem,
which is smooth for $\varepsilon>0$
and uniformly bounded in $\varepsilon\geqslant 0$.
As is well known,
in a similar problem about the propagation of shock waves,
the existence of such an approximation distinguishes a unique
solution.

By $t^{*}\in[0,t_{1})$ we denote the instant of confluence
of free boundaries.
Then, for any $t\leqslant t^{*}-\delta$ for all $\delta>0$,
we see that the limit problems have solutions.
The asymptotic solution of system (\ref{PF:1})
has the form
\begin{align}\label{AS:1}
\theta^{as}_{\varepsilon}&=\bar{\theta}^{-}(x,t)+
\left(\bar{\theta}^{+}(x,t)-\bar{\theta}^{-}(x,t)\right)
\omega_{1}\left(\frac{-x+\hat{\varphi}_2(t)}{\varepsilon}\right)
\omega_{1}\left(\frac{x-\hat{\varphi}_1(t)}{\varepsilon}\right),
\\
\label{AS:2}
u^{as}_{\varepsilon}&=
1+\omega_{0}\left(\frac{-x+\hat{\varphi}_1(t)}{\varepsilon}\right)
\\
&\qquad
+\omega_{0}\left(\frac{x-\hat{\varphi}_2(t)}{\varepsilon}\right)
+\varepsilon\left[\frac{\theta^{as}_{\varepsilon}}{2}+
\omega\left(t,\frac{x-\hat{\varphi}_1(t)}{\varepsilon},
\frac{x-\hat{\varphi}_2(t)}{\varepsilon}\right)\right].\notag
\end{align}
Here $\omega_{1}(z)\to 0,1$ as $z\to\mp\infty$,
$\omega^{(k)}_{1}(z)\in\mathbb{S}(\mathbb{R}^{1}_{z})$ for $k>0$,
$\hat{\varphi}_i(t)$, $i=1,2$, are smooth functions,
$\hat{\varphi}_1\leq\hat{\varphi}_2$,
$\omega_{0}(z)=\tanh(z)$, and $\omega(t,z_{1},z_{2})\in
C^{\infty}([0,t^{*}];\,\mathbb{S}(\mathbb{R}^{2}_{z}))$.
By $\mathbb{S}(\mathbb{R}^n)$ we denote the Schwartz space
of smooth rapidly decreasing functions.
If the initial data for (\ref{PF:1}) has the form
(\ref{AS:1}), (\ref{AS:2}) at $t=0$, then,
for $t\leqslant t^{*}-\delta$, we have the estimate
\begin{equation*}
\|u-u^{as}_{\varepsilon};C(0,T;L^{2}(\mathbb{R}^{1})\|+\|\theta-
\theta^{as}_{\varepsilon};\mathcal{L}^{2}(Q)\|\leqslant
c\varepsilon^{\mu},\qquad \mu\geqslant  3/2,
\end{equation*}
where $(\theta,u)$ is a solution of system (1)  
(see~\cite{1,2}).
Here
$Q=\Omega\times[0,t^{*}-\delta)$, and the constant $c$ is
independent of~$\varepsilon$.

The main obstacle to the construction of solutions
of the form (\ref{AS:1}), (\ref{AS:2}),
which could be used to describe the confluence of free
boundaries, is the fact that,
instead of an ordinary differential equation
whose solution is the function $\omega_{0}(z)$~\cite{12,2},
in the case of confluence of free boundaries, 
we must deal with a partial differential equation
for which the explicit form of the exact solution is unknown.

In the present paper, we use the technique of
the weak asymptotics method \cite{6,11},
which allows us to avoid this problem.
Let us explain several basic points.

\begin{definition}\label{WkMDef}
A family of functions $f(x,t,\varepsilon)\in L^1(Q)$
integrable with respect to~$x$
for all $t\in[0,t_{1}]$ and for $\varepsilon>0$
admits the estimate
$\mathcal{O}_{\mathcal{D}'}(\varepsilon^\nu)$
if, for any test function
$\zeta(x)\in C^{\infty}_{0}$,
we have the estimate
\begin{equation}\label{WkMDef:1}
\left|\int_{\Omega}f(x,t,\varepsilon)\zeta(x)dx\right|\leqslant
C_{t_{1},\zeta}\varepsilon^{\nu},
\end{equation}
where the constant $C_{t_{1},\zeta}$
depends on~$t_{1}$ and the test function $\zeta(x)$.
\end{definition}

Generalizing (\ref{WkMDef:1}),
we shall say that the family of distributions
$f(x,t,\varepsilon)$ depending on~$t$ and~$\varepsilon$
as on parameters admits the estimate
$\mathcal{O}_{\mathcal{D}'}(\varepsilon^\nu)$
if, for any test function $\zeta(x)$, we have the estimate
\begin{equation*}
\langle f(x,t,\varepsilon), \zeta(x)\rangle
=\mathcal{O}(\varepsilon^\nu),\quad
0\leqslant t \leqslant t_{1}.
\end{equation*}

\begin{example}
Let $\omega(z)\in \mathbb{S}(\mathbb{R}^{1})$,
$\int_{\mathbb{R}^{1}}\omega(z)\,dz=1$,
and let $x_{0}\in[l_{1},l_{2}]$.
Then we have
\begin{equation*}
\omega\left(\frac{x}{\varepsilon}\right)-\delta(x)
=\mathcal{O}_{\mathcal{D}'}(\varepsilon),
\qquad
\omega\left(\frac{x-x_{0}}{\varepsilon}\right)
=\delta(x-x_{0})\int_{\mathbb{R}^{1}}\omega(z)\,dz
+\mathcal{O}_{\mathcal{D}'}(\varepsilon).
\end{equation*}
\end{example}

\begin{example}
Suppose that $\omega_{i}(z) \in C^{\infty}$, \
$\left(\omega_{i}\right)'\in\mathbb{S}(\mathbb{R}^{1})$, \
$\lim_{z\to-\infty}\omega_{i}(z)=0$, \
and 
$\lim_{z\to+\infty}\omega_{i}(z)=1$, $i=1,2$.
Then we have
\begin{description}
\item{(a)}
\begin{equation*}
\omega_{i}\left(\frac{x-x_{i}}{\varepsilon}\right)-H(x-x_{i})
=\mathcal{O}_{\mathcal{D}'}(\varepsilon).
\end{equation*}
\item{(b)}
\begin{align*}
&\omega_{1}\left(\frac{x-x_{1}}{\varepsilon}\right)
\omega_{2}\left(\frac{x-x_{2}}{\varepsilon}\right)\\
&\qquad =B_{1}\left(\frac{\triangle x}{\varepsilon}\right)H(x-x_{1})
+B_{2}\left(\frac{\triangle x}{\varepsilon}\right)H(x-x_{2})
+\mathcal{O}_{\mathcal{D}'}(\varepsilon),
\end{align*}
\end{description}
where $\triangle x=x_{1}-x_{2}$, \ $B_{i}(\rho)\in C^{\infty}$, \
$B_{i}'(\rho)\in\mathbb{S}(\mathbb{R}^{1})$, \
$B_{1}(\rho)+B_{2}(\rho)=1$, \ $B_{1}(+\infty)=1$, and
$B_{1}(-\infty)=0$.
\end{example}

\begin{example}
The preceding relations easily imply the formula
\begin{align*}
H(x-x_1)H(x-x_2)
=B\left(\frac{\triangle x}{\varepsilon}\right)H(x-x_1)
+(1-B)H(x-x_2)+\mathcal{O}_{\mathcal{D}'}(\varepsilon),
\end{align*}
where $B(\rho)\in C^\infty$ is the function $B_1$ 
from the preceding example.
\end{example}

\begin{example}
The following relation is a corollary of Definition~1:
$$
\bigg(\frac{d}{dx}\bigg)^m \mathcal{O}_{\mathcal{D}'}(\varepsilon^\alpha)
=\mathcal{O}_{\mathcal{D}'}(\varepsilon^\alpha)
$$
for all $m\in\mathbb{Z}_+$ and $\alpha>0$.
Generally speaking, this is not true for the derivatives
w.r.t.~$t$.
\end{example}

The relations in Examples~2\,(a), 2\,(b) and~4 are obvious,
but, for the reader's convenience,
we prove some of these formulas in Section~5.

Here we only note that,
in view of items~(a) and~(b),
one can say that the product
of approximations of the Heaviside functions
or of the Heaviside functions themselves
is a linear combination
with accuracy up to small terms
($\sim\mathcal{O}_{\mathcal{D}'}(\varepsilon)$).
The general nonlinear functions
of linear combinations of approximations
of the Heaviside functions
can be linearized similarly.
Precisely this property underlies
the constructive study of the interaction
between nonlinear waves with localized fast variations.

\begin{definition}\label{Def:1}
A pair of smooth functions $(\check{u}, \check{\theta})$
is a weak asymptotic solution of the phase field system
(\ref{PF:1}) if, for any test functions
$\zeta(x),\xi(x)\in C^{\infty}_{0}(\Omega)$,
the following relations hold:
\begin{align}\label{Defin:1}
&\int_{\Omega}\left(\check{u}_{t}+\check{\theta}_{t}\right)\zeta\,dx
+\int_{\Omega}\check{\theta}_{x}\zeta_{x}\,dx
= \mathcal{O}(\varepsilon),
\\
\label{Defin:2}
&\varepsilon\int_{\Omega} \check{u}_{t}\check{u}_{x}\xi\,dx
+\frac{\varepsilon}{2}\int_{\Omega}\check{u}^{2}_{x} \xi_{x}\,dx
-\frac{1}{\varepsilon}\int_{\Omega}
\left(\frac{\check{u}^{4}}{4}
-\frac{\check{u}^{2}}{2}+\frac{1}{4}\right)\xi_{x}\,dx
\\
&\qquad
+\varkappa\int_{\Omega}\check{u}
\frac{\partial}{\partial x}(\check{\theta}\xi)\,dx
= \mathcal{O}(\varepsilon^\mu), \qquad \mu\in(0,1/2).\notag
\end{align}
\end{definition}

The left-hand side of Eq.~(\ref{Defin:2}) is obtained by
multiplying the second 
equation in system (\ref{PF:1}) by $\check{u}_{x}$ and integrating
by parts. The reminders $\mathcal{O}(\varepsilon^\alpha)$,
$\alpha=1,\mu$, in the right-hand sides of (\ref{Defin:1}) and
(\ref{Defin:2}) must be bounded locally in $t$, i.e.,
for $t\in[0,t_{1}]$, we have
\begin{equation*}
\max_{0\leqslant t\leqslant
t_{1}}\left|\mathcal{O}(\varepsilon)\right|\leqslant
C_{t_{1}}\varepsilon,\qquad C_{t_{1}}=\mathrm{const}.
\end{equation*}
This construction was introduced and analyzed in~\cite{4}.

The fact that a noninteger exponent appears in the right-hand
side of~(\ref{Defin:2}) is not directly related
to the technique of the weak asymptotics method,
see Examples 1--4. The source of this noninteger exponent
is the nonsmoothness of the function $\theta$ (the temperature),
which appears at the instant of confluence of the free boundaries.

Relations (5)--(6) can be rewritten as
\begin{align*}
\check{\theta}_t+\check{u}_t
&=\frac{\partial^2\check{\theta}}{\partial x^2}
+\mathcal{O}_{\mathcal{D}'}(\varepsilon),
\\
\varepsilon\check{u}_t
&=\varepsilon\frac{\partial^2\check{u}}{\partial x^2}
+\check{u}(1-\check{u})+\mathcal{O}_{\mathcal{D}'}(\varepsilon^\mu).
\end{align*}

The paper is organized as follows.

In Section~2, we explain the structure of ansatzes
of approximations of the temperature and the order function.
These ansatzes are constructed under the assumption
the classical sharp fronted solution of the Stefan problem
exists. At the beginning of Section~2,
we formulate what we exactly need.

Next, in Section~3, we substitute the constructed ansatzes
in system~(1) and derive equations for the unknown functions
contained in the ansatzes.

The results of these calculations are summarized in
Theorems~1--3 in Section~3. The final result of this paper
is formulated in Theorem~4 in Section~3. It states that
the assumption on the existence of the classical solution
is sufficient for constructing formulas for the weak asymptotic
solution of system~(1) in the sense of Definition~2.
Next, in Section~4, we analyze the constructed formulas and
derive the following effects:
\begin{description}
\item{(a)}
the weak asymptotic solution is smooth for $t>t^*$,
the absolute values of the free boundaries velocities
are equal to each other at the contact moment;
\item{(b)}
the temperature has a negative jump at the instant 
and at the point of confluence of the free boundaries,
and this jump is equal to the half-sum
of the limits of the velocities of the free boundaries 
as $t\to t^*-0$.
\end{description}

In particular, it follows from (a) and (b) that 
the velocities of the free boundaries have jumps at the point 
of contact.

These effects can also be discovered in the numerical analysis
of the process of confluence of free boundaries.
Some helpful technical results are given in Section~5.
We note that the only example known to the authors,
where the confluence of the free boundaries is studied,
is given in \cite{13}.

In general, this paper turned out to be 
technically rather complicated and long in spite of the fact
that the details in several justifications in Section~5 
were omitted.

Although, as was shown above,  
the results obtained here are a necessary step 
in the study of the multidimensional problem,
this paper shows that the following classification 
can be introduced:

(1) problems in which the confluence of free boundaries
leads to disappearance of one of the phases;

(2) problems in which the domain occupied by one of the phases 
changes its connectivity, but the number of phases remains 
the same.

In this paper, we consider an example precisely 
from the first class of problems.
As was mentioned above, we do not justify the asymptotics
of the constructed solution. 
But this can be done based on 
our constructions. It is easy to see that the main role here 
is played by the (not proved) existence of the classical
solution up to the moment of confluence of the free boundaries. 
This assumption reduces justifying 
the constructed weak asymptotic solution 
to estimating the soluton of the heat equation 
with the right-hand side $f_\varepsilon$  
admitting the estimate 
$$
f_\varepsilon=O_{\mathcal{D}'}(\varepsilon^\mu),\qquad 
\mu\in(0,1/2)
$$
and with zero initial and boundary conditions.
An analysis of the structure of this right-hand side shows that
$f_\varepsilon\cdot\varepsilon^{-\mu}$ as $\varepsilon\to0$ is a
linear combination of functions $\delta'(x-\varphi_i)$ 
and $\delta(x-\varphi_i)$, $i=1,2$,  
with coefficients depending on $t,\tau$, and these coefficients
converge fast to zero as $\tau\to\pm\infty$. 

Hence we can conclude that for $x\ne\varphi_i$ 
the solution of this heat equation belongs 
to $C^\infty$ for $\varepsilon\geq0$ and admits
the estimate $\mathcal{O}(\varepsilon^\mu)$. 
In the whole domain $\Omega\times[0,T]$, 
a rough analysis based on general theorems~\cite{14}
shows that the solution belongs to $W^{-\delta}_2$, $\delta>0$,
and has the estimate $\mathcal{O}(\varepsilon^\mu)$ in the norm
of this space.  In particular, the weak limit 
of the constructed weak asymptotic solution 
is equal to the exact generalized global solution 
of the heat equation in the phase field system.

\section{Ansatz of the approximation of the solution
of the Stefan problem with kinetic undercooling}\label{S:Constr}

We recall that, in our case,
the Stefan problem with kinetic undercooling
has the form
\begin{equation}\label{SH:1}
\frac{\partial\overline{\theta}}{\partial t}
=\frac{\partial^{2}\overline{\theta}}{\partial x^{2}},\qquad
x\in[l_{1},l_{2}],\quad x\neq \hat{\varphi}_{i},\quad i=1,2,
\end{equation}
\begin{align}\label{SGT:1}
\left.\overline{\theta}\right|_{x=\hat{\varphi}_{i}}
&= (-1)^{i+1} 2^{-1}\hat{\varphi}_{it},
\\
\label{SSt:1}
\left.\left[\frac{\partial\overline{\theta}}{\partial x}\right]
\right|_{x=\hat{\varphi}_{i}}
&= 2(-1)^{i+1}\hat{\varphi}_{it}.
\end{align}
Relations (\ref{SH:1})--(\ref{SSt:1}) are supplemented
with (Dirichlet or Neumann) boundary conditions
for $x=l_{i}$, $i=1,2$, and
with a consistent initial condition.
Obviously, problem (\ref{SH:1})--(\ref{SSt:1})
can be written as
\begin{align}\label{SH:R}
L\overline{\theta}&=2\hat{\varphi}_{2t}
\delta(x-\hat{\varphi}_{2}) - 2\hat{\varphi}_{1t}
\delta(x-\hat{\varphi}_{1}),
\\
\qquad\left.\overline{\theta}\right|_{x=\hat{\varphi}_{i}}
&=(-1)^{i+1} 2^{-1} \hat{\varphi}_{it}.\notag
\end{align}
We shall assume that the initial conditions
are chosen so that the problem in question has
the classical solution,
$\hat{\varphi}_{1}(0)<\hat{\varphi}_{2}(0)$,
and there exists a $t=t^{*}\in[0,t_{1}]$ such that
$\hat{\varphi}_{1}(t^{*}-0)=\hat{\varphi}_{2}(t^{*}-0)$.
More precisely, we assume that
\begin{description}
\item{(i)}
the limits $\lim_{t\to t^{*}-0}\hat{\varphi}_{it}$ exist, $i=1,2$;
\item{(ii)}
\begin{align*}
\overline{\theta}(x,t)&\in C^{1}\left([0,t^{*}),
C^{2}\left(\overset{\circ}{\Omega}{}^{-}(t)\cup
\overset{\circ}{\Omega}{}^{+}_{l}(t)
\cup\overset{\circ}{\Omega}{}^{+}_{r}(t)\right)\right)\\
&\cap\left(C^{1}\left(\Omega^{-}(t)\right)\cup
C^{1}\left(\Omega^{+}_{l}(t)\right)\cup
C^{1}\left(\Omega^{+}_{r}(t)\right)\right)\cap C(\Omega)
\end{align*}
(here $\overset{\circ}{D}$ denotes the interior of the domain~$D$),
cf.~\cite{9}.
\end{description}

As will be shown below, these assumptions permit describing
the interaction (the confluence of free boundaries)
constructively.

We shall construct this approximation (ansatz)
as a weak asymptotic solution of system~(1).
First, we introduce the ansatz of the order function.
It has the form
\begin{align}\label{AU:1}
 \check{u}&=\frac{1}{2}\left[1+
 \omega_{0}\left(\beta\frac{\varphi_{1}-x}{\varepsilon}\right)+
 \omega_{0}\left(\beta\frac{x-\varphi_{2}}{\varepsilon}\right)\right.\\
          &   -\left.
  \omega_{0}\left(\beta\frac{\varphi_{1}-x}{\varepsilon}\right)
  \omega_{0}\left(\beta\frac{x-\varphi_{2}}{\varepsilon}\right)\right]\notag.
\end{align}

Here the unknowns are the functions $\beta$ and
$\varphi_{i}=\varphi_{i}(t,\varepsilon)$, $i=1,2$.
In what follows, we write them more precisely.
Now we only note that if $\beta>0$,
then, for $\varphi_{1}<\varphi_{2}$,
the function $\check{u}$ coincides
up to $\mathcal{O}(\varepsilon^{N})$
with the sum of the first three terms
in the right-hand side of (\ref{AS:2}),
for $\varphi_{1}=\varphi_{2}$,
we have
$\check{u}=1+\mathcal{O}_{\mathcal{D}'}(\varepsilon)$,
and for $\varphi_{1}>\varphi_{2}$,
we have
$\check{u}=1+\mathcal{O}(\varepsilon^N)$.
These relations can easily be verified directly.
We note that, under the above assumptions,
we can continue the functions
$\hat{\varphi}_{i}(t)$ to the interval $[0,t_{1}]$
preserving the smoothness and the sign of the derivatives.
We choose such continuations and denote them
by $\varphi_{i0}(t)$.
Now, we can write the functions
$\varphi_{i}(t,\varepsilon)$ and
$\beta(t,\varepsilon)$ more precisely.
Namely, we set
\begin{equation}\label{PHI:1}
\varphi_{i}(t,\varepsilon)
=\varphi_{i0}(t)+\psi_{0}(t)\varphi_{i1}(\tau,t),
\end{equation}
where
\begin{equation}\label{PhiTau:1}
\psi_{0}(t)=\varphi_{20}(t)-\varphi_{10}(t),\quad
\tau=\psi_{0}/\varepsilon.
\end{equation}
We note that
\begin{align*}
&\tau\to \infty,\quad t<t^{*},\\
&\tau\to -\infty,\quad t>t^{*},
\end{align*}
as $\varepsilon\to 0$.
Thus, the value of the variable~$\tau$ characterizes
the process of confluence of free boundaries. Indeed, we could
set $\tau=(t^{*}-t)/\varepsilon$, but, for (\ref{PhiTau:1}),
the next formulas become simpler
(this can be seen from the formulas in Example~2 if we put
$x_i=\varphi_{i0}(t)$).

Similarly, we set
\begin{equation}\label{BETADef:1}
\beta(\tau)=1+\beta_{1}(\tau).
\end{equation}
In addition, we assume that
\begin{align}
&\beta^{(\alpha)}_{1}(\tau),\ \varphi_{i1}^{(\alpha)}(\tau)\to
\mathcal{O}(\tau^{-1}),\quad \tau\to\infty, \quad
\alpha=0,1.\label{BETAPHI:1}\\
&0<\delta_{1}<\beta_{1}(\tau)<\delta_{2},\quad \delta_{1},\
\delta_{2}=\mathrm{const},\notag\\ &\varphi_{i1}\to
\varphi^{-}_{i1},\quad \varphi_{i1}-\varphi^{-}_{i1}
=\mathcal{O}(|\tau|^{-1}),\quad\tau\to -\infty,\quad
\varphi^{-}_{i1}=\mathrm{const}.\notag
\end{align}
Assumptions (\ref{BETAPHI:1}) for the functions $\varphi_{i1}$
are the assumption that as $\varepsilon\to 0$
the functions $\varphi_{i}(t,\varepsilon)$
approximate continuous functions with a possible discontinuity
(jump) of the derivatives at $t=t^{*}$.

More precisely, the functions $\varphi_i(t,\varepsilon)$
determined in (12) are approximations of the functions
$$
\varphi_i(t,0)=\varphi_{i0}+\psi_0 \varphi^-_{i1} H(-\psi_0),
$$
where $\varphi^-_{i1}=\lim_{\tau\to-\infty}\varphi_{i1}(\tau,t)$,
$\varphi_i(t,\varepsilon)-\varphi_i(t,0)=\mathcal{O}(\varepsilon)$,
and $H(z)$ is the Heaviside function.

Now we describe the ansatz of the approximation
of the temperature $\bar{\theta}(x,t)$.
Here we also must use the continuation procedure
for matching the temperature for $t<t^{*}$,
when there are three subdomains of the domain~$Q$,
and the temperature for $t>t^{*}$,
when there is only one phase.
For this,
we first introduce a ``model'' of the temperature
whose continuation is reduced to the problem of
continuation of functions depending only on the time~$t$.
This model must have the same structure as the temperature
$\bar{\theta}$. Obviously, the simplest function
of this form is a function linear in~$x$
in the domains $\Omega^{+}_{l,r}(t)$
and quadratic in the domain $\Omega^{-}(t)$.

Namely, we set
\begin{align*}
&\left.\frac{\partial \bar{\theta}}{\partial x}
\right|_{x=\hat{\varphi}_{2}\pm0}=\pm\gamma^{\pm}_{2},
\\
&\left.\frac{\partial \bar{\theta}}{\partial x}
\right|_{x=\hat{\varphi}_{1}\pm0}=\pm\gamma^{\pm}_{1}, 
\end{align*}

We note that, by assumptions~(i) and~(ii),
the functions $\gamma_{j}^{\pm}$, $j=1,2$,
depend on~$t$, are continuous for $t<t^{*}$,
and have limits as $t\to t^{*}-0$.
Therefore, they can be continued to the interval
$[0,t_{1}]$, $t_{1}>t^{*}$,
so that the properties of being continuous
and the signs are preserved.
We shall use the previous notation for the continued functions.
Now we can introduce the temperature model
\begin{align}\label{ModT:1}
\check{T}&=\gamma_{1}^{-}(\varphi_{1}-x)H(\varphi_{1}-x)
+\gamma_{2}^{+}(x-\varphi_{2})H(x-\varphi_{2})\notag\\
&+\gamma^{-}(x,t)\frac{(\varphi_{1}-x)(x-\varphi_{2})}{\psi}
H(x-\varphi_{1})H(\varphi_{2}-x)\\
&+\hat{\gamma}(x,t)\frac{(\varphi_{1}-x)(x-\varphi_{2})}{\psi}
H(\varphi_{1}-x)H(x-\varphi_{2})+I,\notag
\end{align}
where $\gamma^{-}$, $\hat{\gamma}$, and~$I$
are linear functions of~$x$,
$\left.\gamma^{-}\right|_{x=\varphi_{i}}=\gamma^{-}_{i}(t)$,
$\left.\hat{\gamma}\right|_{x=\varphi_{i}}=\hat{\gamma}_{i}(t)$,
and $\left.I\right|_{x=\varphi_{i}}=(-1)^{i+1}\varphi_{it}$.
In more detail, the functions
$\hat{\gamma}(x,t)$ and $\gamma^{-}(x,t)$ can be written
in the form
\begin{align*}
&\gamma^{-}=\frac{\gamma_{1}^{+}-\gamma_{2}^{-}}{2}
-(x-x^{*})\frac{\gamma_{1}^{+}+\gamma_{2}^{-}}{\psi}
\\
&\hat{\gamma}=\frac{\hat{\gamma}_{1}+\hat{\gamma}_{2}}{2}
-(x-x^{*})\frac{\hat{\gamma}_{1}-\hat{\gamma}_{2}}{\psi},
\qquad x^{*}=\frac{\varphi_{1}+\varphi_{2}}{2}.
\\
&I=\frac{\varphi_{1t}-\varphi_{2t}}{2}
-(x-x^*)\frac{\varphi_{1t}+\varphi_{2t}}{2}.
\end{align*}
We note that, by Lemma~3 in Section~5,
for $t\leqslant t^{*}$, we have the relations
\begin{equation}\label{OPhiPhi}
\varphi_{i}(t,\varepsilon)=\hat{\varphi}_{i}(t)
+\mathcal{O}(\varepsilon), \qquad i=1,2.
\end{equation}
This implies that, for $t<t^{*}$,
\begin{align}\label{TDelta}
&\left.\left[\frac{\partial \check{T}}{\partial x}\right]
\right|_{x=\varphi_{i}}=
\left.\left[\frac{\partial \bar{\theta}}{\partial x}\right]
\right|_{x=\hat{\varphi}_{i}},\\
&\left[\frac{\partial \check{T}}{\partial x}\right]\delta(x-\varphi_{1})
=\left[\frac{\partial \bar{\theta}}{\partial x}\right]
\delta(x-\hat{\varphi}_{1})+\mathcal{O}_{\mathcal{D}'}(\varepsilon).\notag
\end{align}
Using the formula in Example~3 above,
we rewrite the product
\begin{equation*}
H(x-\varphi_{1})H(\varphi_{2}-x)
\end{equation*}
as follows
\begin{gather}\label{ProdH:1}
H(x-\varphi_{1})H(\varphi_{2}-x)
=B(\rho)[H(x-\varphi_{1})-H(x-\varphi_{2})]
+\mathcal{O}_{\mathcal{D}'}(\varepsilon).
\\
H(\varphi_{1}-x)H(x-\varphi_{2})
=(1-B(\rho))[H(x-\varphi_{2})-H(x-\varphi_{1})]
+\mathcal{O}_{\mathcal{D}'}(\varepsilon).
\notag
\end{gather}
Here $B(\rho)\in C^\infty$, $B(\rho)\to1$, $\rho\to\infty$,  
$B(\rho)\to0$, $\rho\to-\infty$, 
$B^{(\alpha)}=\mathcal{O}(|\rho|^{-N})$, 
$|\rho|\to\infty$ for any $N>0$, $\alpha>0$.

In what follows, we write $B(\tau)$ instead of $B(\rho)$. 
The difference is that $B(\rho)$ is an unknonw function 
($\rho$ is unknown), while $B(\tau)$ is a known function.
The replacement of $B(\rho)$ by $B(\tau)$ in (19) and (20)
implies a correction of order 
$\mathcal{O}_{\mathcal{D}'}(\varepsilon)$ 
in the right-hand sides.

But we do not use this remark and, by definition, we set
\begin{align}
\check{T}&=\gamma^-_1(\varphi_1-x)H(\varphi_1-x)
+\gamma^-_2(x-\varphi_2)H(x-\varphi_2)
\\
&\qquad 
+\gamma^-(x,t)\frac{\varphi_1-x)(x-\varphi_2)}{\psi}
B(\tau)[H(x-\varphi_1)-H(x-\varphi_2)]
\notag\\
&\qquad 
+\hat\gamma(x,t)\frac{\varphi_1-x)(x-\varphi_2)}{\psi}
(1-B(\tau))[H(x-\varphi_2)-H(x-\varphi_1)]
+I.
\notag
\end{align}

Now we can write the expression
$\frac{\partial^{2}\check{T}}{\partial x^{2}}$
uniformly in $\psi=\varphi_{2}-\varphi_{1}$.
We have
\begin{align}\label{D2T}
&\frac{\partial^{2}\check{T}}{\partial x^{2}}
=F_{1}(x,t,\tau)
\\
&\qquad
+\frac{\partial^2}{\partial x^2}
\bigg(\gamma^-\frac{(\varphi_1-x)(x-\varphi_2)}{\psi}\bigg)
B(\tau)[H(x-\varphi_{1})-H(x-\varphi_{2})]\notag
\notag\\
&\qquad
+\frac{\partial^2}{\partial x^2}
\bigg(\hat\gamma\frac{(\varphi_1-x)(x-\varphi_2)}{\psi}\bigg)
(1-B(\tau)) [H(x-\varphi_{2})-H(x-\varphi_{1})]
\notag\\
&\qquad+\left[\gamma_{2}^{+}+\gamma_{2}^{-}B(\tau)
+\hat{\gamma}_{2}(1-B(\tau))\right]\delta(x-\varphi_{2})
\notag\\
&\qquad+\left[\gamma_{1}^{-}+\gamma_{1}^{+}B(\tau)
-\hat{\gamma}_{1}(1-B(\tau))\right]\delta(x-\varphi_{1})
+\mathcal{O}_{\mathcal{D}'}(\varepsilon),
\notag
\end{align}
where $F_{1}(x,t,\tau)$ is a bounded piecewise continuous
function.
We set
$\gamma_{2}^{+}=-\hat{\gamma}_{2}$ and
$\gamma_{1}^{-}=\hat{\gamma}_{1}$
and see that the coefficients for $\delta(x-\varphi_{i})$ in
(\ref{D2T}) take the form
\begin{align}\label{CoefDelta}
&\left[\gamma_{2}^{-}+\gamma_{2}^{+}\right]B(\tau)\delta(x-\varphi_{2})
+\left[\gamma_{1}^{-}+\gamma_{1}^{+}\right]B(\tau)\delta(x-\varphi_{1}).
\end{align}
Now we note that the inequalities
\begin{align*}\label{HHPsi}
\left[H(x-\varphi_{1})-H(x-\varphi_{2})\right]/\psi >0,\qquad
\left[H(x-\varphi_{2})-H(x-\varphi_{1})\right]/\psi <0,
\end{align*}
hold for $\psi\neq 0$.
We recall that, by (\ref{TDelta}) and the Stefan condition
(\ref{SSt:1}),
\begin{equation*}
\varphi_{10t}=\gamma_{1}^{+}+\gamma_{1}^{-},\qquad
\varphi_{20t}=-(\gamma_{2}^{+}+\gamma_{2}^{-})
\end{equation*}
for $t<t^*$ and assume that the continuations of the functions
contained in these relations are chosen so that these relations
hold for $t\in [0,t_1]$.

The assumption that $\overline{\theta}(x,t)$ 
is the classical solution
of problem (\ref{SH:1})--(\ref{SSt:1}) implies that
\begin{equation} 
\bar{\theta}-\check{T}
=\check{q}\in C([0,t^{*});C^{1}(\Omega))
\end{equation}
and the limit $\check{q}$ exists as $t\to t^{*}-0$ and for
$x\in\Omega$. Therefore, we can continue the function
$\check{q}$ to the domain $\Omega\times[0,t_{1}]$, where
$t_{1}>t^{*}$ and the properties of smoothness are preserved,
and moreover,
\begin{equation*}
\left.\check{q}\right|_{x=\hat{\varphi}_{i}}=0, \qquad i=1,2,
\qquad t<t^{*}.
\end{equation*}
We shall construct the global temperature $\check{\theta}$
in the form
\begin{equation}\label{CheckTh}
\check{\theta}=e(x)\check{T}+\check{q}+\hat{q},
\end{equation}
where $\hat{q}$ is the desired function,
$e(x)\in C_{0}^{\infty}([l_{1},l_{2}])$,
and $e\equiv 1$ for
$x\in[\hat{\varphi}_{1}(0),\hat{\varphi}_{1}(0)]$.

\section{Construction of the weak asymptotic
solution}\label{Constr:1}

We fist consider the heat equation in system (\ref{PF:1}).
Using the formulas of the weak asymptotics method
(see Section~5) and taking (\ref{AU:1}) into account,
we obtain
\begin{align}\label{Udivt:1}
\frac{\partial \check{u}}{\partial
t}&=\left[\frac{\varphi_{1t}}{2}(2-B_{\dot{0}0})+
\frac{\beta_{\tau}(\varphi_{20t}-\varphi_{10t})}{2\beta^{2}}
B^{z}_{\dot{0}0}\right]\delta(x-\varphi_{1})\\
&+\left[-\frac{\varphi_{2t}}{2}(2-B_{\dot{0}0})+
\frac{\beta_{\tau}(\varphi_{20t}-\varphi_{10t})}{2\beta^{2}}
B^{z}_{\dot{0}0}\right]\delta(x-\varphi_{2})+
\mathcal{O}_{\mathcal{D}'}(\varepsilon),\notag
\end{align}
where the estimate $\mathcal{O}_{\mathcal{D}'}(\varepsilon)$
is uniform in $\tau$,
\begin{align*}
B_{\dot{0}0}&=\int_{\mathbb{R}^{1}}\dot{\omega}_{0}(z)
\omega_{0}(-\eta-z)\,dz,
\qquad
B^{z}_{\dot{0}0}=\int_{\mathbb{R}^{1}}
z\dot{\omega}_{0}(z)\omega_{0}(-\eta-z)\,dz,\\
\eta&=\rho\beta,
\end{align*}
and the function $\rho(\tau)$ is determined by the formula
\begin{equation}\label{RhoDef:1}
\rho(\tau)=\frac{\varphi_{2}-\varphi_{1}}{\varepsilon}.
\end{equation}

We assume that function $\check{q}$ and its continuation are
chosen so that the function $\check{q}$ satisfies the boundary
conditions of the original problem.
Then the boundary conditions for the function $\hat{q}$ are
zero.

We substitute the function $\check{\theta}$
determined by relation (24) and expression (25) 
for $\partial\check{u}/\partial t$ into Eq.~(5).
With accuracy $\mathcal{O}_{\mathcal{D}'}(\varepsilon)$,
we obtain
\begin{equation*}
L\check{\theta}+\frac{\partial \check{u}}{\partial t}
=Le\check{T}+L{q}+A_1\delta(x-\varphi_1)
+A_2\delta(x-\varphi_2),
\end{equation*}
where
$$
q=\check{q}+\hat{q},\qquad 
A_i=(-1)^{i+1}\frac{\varphi_{it}}{2}(2-B_{00})
+\frac{\beta_\tau\psi'_{ot}}{2\beta^2}B^z_{00},\qquad
i=1,2.
$$
We let $\tau$ tend to $\infty$ (i.e., for $t<t^*$)
and, in view of (10) and (18), obtain
$$
L(e\check{T}+{q})=2\hat{\varphi}_{2t}
\delta(x-\hat{\varphi}_{2}) - 2\hat{\varphi}_{1t}
\delta(x-\hat{\varphi}_{1}).
$$

Since $e\check{T}+\check{q}\to\overline{\theta}$ as
$\tau\to\infty$, we see that $\hat{q}\to0$ as $\tau\to\infty$.
The total equation for the function $q=\check{q}+\hat{q}$ 
has the form
\begin{align}
&L(q+e(x)I) = F(x,t,\tau) 
\\
&\qquad
-\frac{\partial^2}{\partial x^2}
\bigg(\gamma^-\frac{(\varphi_1-x)(x-\varphi_2)}{\psi}\bigg)
B(\tau)[H(x-\varphi_{1})-H(x-\varphi_{2})]\notag
\notag\\
&\qquad
-\frac{\partial^2}{\partial x^2}
\bigg(\hat\gamma\frac{(\varphi_1-x)(x-\varphi_2)}{\psi}\bigg)
(1-B(\tau)) [H(x-\varphi_{2})-H(x-\varphi_{1})],
\quad x\in\Omega,
\notag
\end{align}
where $F(x,t,\tau)$ is a piecewise continuous function
containing of terms that are uniformly bounded in $\varepsilon$
on $\Omega\times[0,t_1]$ (they appear from $F_1(x,t,\tau)$ in
(21) and the derivatives of the function $\check{q}$).

Equation (27) was derived under the assumption that
(see (22))
\begin{align}
\sum^{2}_{i=1}\{B[\gamma^+_i+\gamma^-_i]-A_i\}\delta(x-\varphi_i)
=\mathcal{O}_{\mathcal{D}'}(\varepsilon).
\end{align}
By $\hat{q}_i$, $\hat{q}^*_i$ 
we denote the typical terms in $q$ such that
\begin{align}
L\hat{q}_1&=-\frac{2(\gamma^+_1+\gamma^-_2)}{\psi}B(\tau)
[H(x-\varphi_2)-H(x-\varphi_1)],
\\
L\hat{q}_2&=\frac{2(\gamma^-_1+\gamma^+_2)}{\psi}(1-B)
[H(x-\varphi_2)-H(x-\varphi_1)],
\notag\\
L\hat{q}^*_1&=2(x-x^*)(\gamma^+_1-\gamma^-_2)\psi^{-2} B(\tau)
[H(x-\varphi_2)-H(x-\varphi_1)],
\notag\\
L\hat{q}^*_2&=-2(x-x^*)(\gamma^-_1-\gamma^+_2)\psi^{-2} B(\tau)
[H(x-\varphi_2)-H(x-\varphi_1)].\notag
\end{align}

With accuracy up to functions smooth in $\Omega$,
we can calculate them using the fundamental solution
of the heat equation (in what follows, we shall consider
only the first equation),
\begin{align}
\hat{q}_1&=-\frac{1}{2\sqrt{2\pi}}\int^{t}_{0}
\frac{(\gamma^+_1(t')+\gamma^-_2(t'))}{\sqrt{t-t'}}
B(\tau(t'))\\
&\times
\int^{\varphi_2}_{\varphi_1}
\frac{ \exp\{-(x-\xi)^2/(t-t')\} }{\psi}\,dt' d\xi.\notag
\end{align}
It is clear that the functions $\hat{q}_i$
are uniformly bounded.
It is also clear that $\hat{q}_i\not\in C^1(\Omega)$, but
$\hat{q}_i\in C^{1,2}(\Omega\times[0,t_1])\setminus
(\{x=\varphi_1\}\cup\{x-\varphi_2\})$.
Therefore, to justify Eq.~(27), we must verify
that no $\delta$-functions arise in calculating 
the derivatives 
$\frac{\partial^2 \hat{q}_i}{\partial x^2}$, 
$\frac{\partial^2 \hat{q}^*_i}{\partial x^2}$ 
and
$\frac{\partial \hat{q}_i}{\partial t}$,
$\frac{\partial \hat{q}^*_i}{\partial t}$.

For this, we note that, for any test function $\zeta(x)$
up to functions smooth in $\Omega$, omitting the number factor,  
we have the relation
\begin{equation}
\langle \hat{q}_i,\zeta(x)\rangle
=\int^{t}_{0}\gamma^-_1(t')
B\int^{\varphi_2}_{\varphi_1}\psi^{-1}
f(\xi,t-t')\,dt]\,d\xi,
\end{equation}
where $f(\xi,t-t')
=\int_{\mathbb{R}^{1}}
\frac{\zeta(x)\exp\{-(x-\xi)^2/(t-t')}{\sqrt{2\pi(t-t')}}$
is the solution of the heat equation $f_t-f_{\xi\xi}=\zeta(\xi)$
at the point $t-t'$.
It is clear that $f(\xi,t)\in C^\infty(\Omega)$ for all~$t$.
Calculating the derivative $\partial^2 \hat{q}_1/\partial x^2$,
we obtain
$$
\Big\langle \frac{\partial^2\hat{q}_i}{\partial x^2},\zeta\Big\rangle
=\langle \hat{q}_1,\zeta''_x\rangle
=\int^{t}_{0}\gamma^-_1(t')B
\int^{\varphi_2}_{\varphi_1}\psi^{-1} f_1(\xi,t-t')\,dt'\,d\xi,
$$
where $f_1(\xi,t)$ is a solution of the equation
$f_t-f_{\xi\xi}=\zeta''(\xi)$.
It is clear that the relation
$$
\Big\langle \frac{\partial^2\hat{q}_i}{\partial x^2},\zeta(x)\Big\rangle
\not= \sum G_i\zeta(\varphi_i)+\mathcal{O}(\varepsilon)
$$
cannot hold for any coefficients $G_i$.
The same is true for $\partial q_i/\partial t$, 
$\partial q^*_i/\partial t$.
Similarly to (31),
with accuracy up to smooth functions,
we can write the functions $\hat{q}^*_i$, $i=1,2$.
They have the same properties as $\hat{q}_i$
with the only additional condition
$$
\hat{q}^*_i|_{x=x^*}=0.
$$
This easily follows from the explicit formula of the type of
(31) and the fact that, for $x=x^*$, the integral with respect
to~$\xi$ is an integral of an odd function over a symmetric
interval.
Therefore, condition (28) is necessary for deriving Eq.~(27).
This implies that the function $\hat{q}+eI$ is uniformly bounded
in $\Omega\times[0,t_1]$ and belongs to~$C^{1,0}$ for
$x\not=\varphi_i$, $i=1,2$.

So if the function $\hat{q}+eI$ satisfies Eq.~(27)
with zero initial conditions and the boundary conditions
that follow from the fact that the function
$\check{\theta}=e\check{T}+\check{q}+\hat{q}$ must satisfy
the boundary conditions of the original problem,
while the functions $\check{T}$ and $\check{q}$
are known
(more precisely, they will be determined after the
functions $\varphi_i$ are found),
then we have the following assertion.

\begin{theorem}\label{TH:1}
Suppose that the function $\check{\theta}$ is determined by
relation {\rm(28)}, and the function
$\hat{q}$ is a solution of Eq.~{\rm(27)}
with zero initial condition and zero boundary conditions.
Suppose that relation {\rm(28)}
holds.

Then the pair of functions $\check{\theta}$ and $\check{u}$
is a weak asymptotic solution of the heat equation
in the phase field system, i.e.,
relation {\rm(5)} holds.
\end{theorem}

\begin{remark}\label{Rem:1}
Expression (28) may seem to be strange, because we did not take
into account the well-known fact that the Dirac functions are
linearly independent. But this fact is taken into account in the
framework of the weak asymptotics method
(see Lemma~2, Section~5).
Here we do not want to mix the substitution of the ansatz
into the equation and the analysis of the results of this
substitution (see Section~4).
\end{remark}

Now we consider the second equation of phase field system,
i.e., the Allen--Cahn equation.
By Definition~\ref{Def:1} (see formula (\ref{Defin:2})),
we must calculate the weak asymptotics of the expression
\begin{equation*}
\mathcal{F}\buildrel\rm def\over=\frac{\partial
\check{u}}{\partial x}\left[\varepsilon
L\check{u}-\frac{\check{u}-\check{u}^{3}}{\varepsilon}
-\varkappa\check{\theta}\right],
\end{equation*}
where the function $\check{\theta}$ is determined by the
relation (20). Calculating similarly as in
Example~2 above (see Section~5 for details),
we obtain
\begin{align}\label{FCal:1}
\mathcal{F}=&V_{1}^{1}\delta(x-\varphi_{1})+V_{2}^{1}\delta(x-\varphi_{2})
\\
&+V_{1}^{2}\delta'(x-\varphi_{1})+V_{2}^{2}\delta'(x-\varphi_{2})
+\mathcal{O}_{\mathcal{D}'}(\varepsilon),\notag
\end{align}
where $V_{j}^{i}$, $i,j=1,2$, are linear combinations of several
convolutions.
Their expressions will be given below.
Here we note the following. It is rather clear that the
expression
\begin{equation*}
\frac{\varepsilon}{2}
\int_{\mathbb{R}^{1}}(\check{u}_{x})^{2}\zeta_{x}\,dx
\end{equation*}
leads in the singular part to functions of the form
$\delta'(x-\varphi_{i})$.

Indeed, according to our construction,
the expression $\varepsilon\check{u}_{x}$
is approximately the function
$(\mathrm{sign}_{\varepsilon}z)'_{z}$, where
$\mathrm{sign}_{\varepsilon}$ is the regularization of the
function $\mathrm{sign}$, i.e., a soliton-type function.
It does not change its structure when squared,
while its derivative divided by $\varepsilon$
becomes the derivative of~$\delta$, i.e.,~$\delta'$.

Similarly, the term
\begin{equation*}
\int_{\mathbb{R}^{1}}\frac{\partial \check{u}}{\partial x}
(\check{u}-\check{u}^{3})\zeta\, dx
\end{equation*}
is transformed as
\begin{equation*}
\int_{\mathbb{R}^{1}}\frac{\partial \check{u}}{\partial x}
(\check{u}-\check{u}^{3})\zeta\,dx
=\int_{\mathbb{R}^{1}}\frac{\partial}{\partial x}
F(\check{u})\zeta\,dx
=-\int_{\mathbb{R}^{1}} F(\check{u})\zeta_{x}\,dx.
\end{equation*}
Similar arguments show that the singular expression itself
in the asymptotics of the last integral
is also of the form $\delta'$.

We denote
\begin{equation}\label{Omega} 
\Omega(z,\eta)=\frac{1}{2}\left\{1+\omega_{0}(z)+\omega_{0}(-z-\eta)
-\omega_{0}(z)\omega_{0}(-z-\eta)\right\}.
\end{equation}

We have the following estimates:
\begin{align}\label{r34} 
\Omega(z,\eta)&=1+f(z,\eta) e^{2\eta}, \qquad \eta\to-\infty,
\\
\Omega(z,\eta)&=\omega_0(z)+f_1(z,\eta) e^{-2\eta}, \qquad \eta\to\infty,
\end{align}
where
$$
\int|f(z,\eta)|\,dz\leq\mathrm{const},\qquad
\int|f_1(z,\eta)|\,dz\leq\mathrm{const}.
$$

These relations readily follow from the explicit form
of the function $\omega_0(z)$,
$$
\omega_0(z)=\frac{e^z-e^{-z}}{e^z+e^{-z}}.
$$
In fact, relations (34) and (35) express
the above-described properties of the ansatz $\check{u}$~(11)
in different terms.

Using the technique of the weak asymptotics method
(see Lemma~7 in Section~5), we can write
the weak asymptotics of the expression
$\int_{\mathbb{R}^1}\zeta\check u_t \check u_x\,dx$
in the form
\begin{align} 
\int_{\mathbb{R}^1}\zeta\check u_t \check u_x\,dx
&=\sum^2_{i=1}\varphi_{it}
\int\Omega'_z(\Omega'_z-\Omega'_\eta)\,dz
\\
&\qquad
+\frac12 \frac{\beta_\tau\psi'_{0t}}{\beta^2}
\int\Omega'_z(\Omega'_z-\Omega'_\eta)\,dz
+\mathcal{O}_{\mathcal{D}'}(\varepsilon).
\notag
\end{align}

Similarly, we obtain (see Lemmas~5 and~7, Section~5):
\begin{align}\label{ThUX} 
\int_{\mathbb{R}^{1}} \theta\check{u}_{x}\zeta\,dx
&=\frac{1}{2}\left(\varphi_{1t}+\left.(\hat{q}+\check{q})
\right|_{x=\varphi_{1}}\right)
\int_{\mathbb{R}^{1}}
\dot{\Omega}_{\eta}(z,\eta)\zeta(\varphi_{1})\,dz\notag\\
&-\frac{1}{2}\left(-\varphi_{2t}+\left.(\hat{q}+\check{q})
\right|_{x=\varphi_{2}}\right)
\int_{\mathbb{R}^{1}}\dot{\Omega}_{\eta}(z,\eta)
\zeta(\varphi_{2})\,dz +\mathcal{O}(\varepsilon^\mu),\quad
\mu\in(0,1/2).
\end{align}

Thus, adding expressions (36) and (37),
we see that the coefficients of the $\delta$-functions
in formula (32) have the form
\begin{equation}
V^1_i=-\bigg[\varphi_{it}B_\Omega
+\frac{\beta_\tau\psi'_{0t}}{\beta^2} B^z_\Omega\bigg]
+(-1)^i(\varphi_{it}+(eI+q)|_{x=\varphi_i})C_\Omega,\quad
i=1,2,
\end{equation}
where
\begin{align}
B_\Omega&=\int\Omega'_z(\Omega'_z-\Omega'_\eta)\,dz,
\notag\\
B^z_\Omega&=\int[z(\Omega'_z-\Omega'_\eta)-(z+\eta)\Omega'_\eta]
(\Omega'_z-\Omega'_\eta)\,dz,
\\
C_\Omega&=\int(\Omega'_z-\Omega'_\eta)\,dz,\qquad
q=\hat{q}+\check{q}.
\notag
\end{align}

Similarly, we obtain (see Lemma~6, Section~5):
\begin{equation}\label{V12:1}
V_{1}^{2}=V_{2}^{2}=\beta\hat{C}-\frac{1}{\beta}\hat{D},
\end{equation}
where
\begin{equation}\label{C:1}
\hat{C}=\frac{1}{4}\int_{\mathbb{R}^{1}}(\Omega'_{z})^{2}\,dz
\end{equation}
\begin{equation}\label{D:1}
\hat{D}=\frac{1}{2}\int_{\mathbb{R}^{1}} F(\Omega)\,dz
\end{equation}

Thus, to obtain $\beta=\beta(\eta)$
we have the equation
(obviously, this is a necessary condition for the relation
$\mathcal{F}=\mathcal{O}_{\mathcal{D}'}(\varepsilon)$ to hold,
see~(\ref{FCal:1})):
\begin{equation}\label{BETA:1}
\beta^{2}=\frac{\hat{D}}{\hat{C}}
\end{equation}
According to (41) and (42),
the right-hand side of the relation (42)
is positive.

So we have proved the following assertion.

\begin{theorem}\label{TH:2}
Suppose that the assumptions of Theorem~{\rm1} and {\rm(43)}
are satisfied and
\begin{equation}\label{SumVDelta:1}
\sum^{2}_{i=1}V^1_1\delta(x-\varphi_1)
+V^1_2\delta(x-\varphi_2)=O_{D'}(\varepsilon).
\end{equation}

Then the pair of functions $\check\theta$, $\check u$
is a weak asymptotic solution of the phase field system
{\rm(\ref{PF:1})} in the sense of Definition~{\rm2}.
\end{theorem}

Thus, under the assumption that the classical solution
of the phase field system exists
(see~(i) and~(ii) above),
relations (27), (28), (43), and (44)
are sufficient conditions for constructing
a weak asymptotic solution of system (1).
In what follows, we prove that the relations mentioned above
are equations for determining the functions
$\beta=\beta(\tau)$ and $\varphi_{i1}(\tau)$, $i=1,2$.
We present an algorithm for solving these equations.

\section{An analysis of the relations obtained}

We begin with relation (28) and, for a while, forget everything
said about the notion of linear independence.

Then, for this relation to hold, it suffices to have the two
relations
\begin{equation}
B[\gamma^+_i+\gamma^-_i]=A_i,\qquad i=1,2.
\end{equation}
In view of (18), we have $\gamma^+_i+\gamma^-_i
=\left.\left[\frac{\partial\overline{\theta}}{\partial x}\right]
\right|_{x=\hat{\varphi}_i(t)}$.
Taking this into account and adding relations (45),
we obtain
\begin{align}
&\psi'_0\left(\rho_r-\frac12\int_{\mathbb{R}^{1}}
\dot\omega(z)\omega_0(-z-\eta)
\left(\rho_\tau-2\frac{\beta_\tau}{\beta}z\right)\right)\,dz\\
&\qquad
=B(\tau)\left\{
\left.\left[\frac{\partial\overline{\theta}}{\partial x}\right]
\right|_{x=\hat{\varphi}_1(t)}
+\left.\left[\frac{\partial\overline{\theta}}{\partial x}\right]
\right|_{x=\hat{\varphi}_2(t)}
\right\} =2B\psi'_{0t}.\notag
\end{align}
Here we used the relation
$$
\psi_t=\varphi_{2t}-\varphi_{1t}
=\psi'_0\rho_\tau+\psi_0(\varphi_{21}-\varphi_{11})_t,
$$
which follows from the definition of the function $\rho$,
see~(25).

We set the last term to be zero,
since we show below that $(\varphi_{21}-\varphi_{11})
=\mathcal{O}(|\tau|^{-1})$.
In view of Lemma~3 in Section~5, in this case we have
$(\varphi_{21}-\varphi_{11})_t\psi_0=\mathcal{O}(\varepsilon)$.
Moreover,
in view of the Stefan conditions and the choice of the
continuation of the functions $\gamma^\pm_i$ and
$\varphi_{0i}(t)$, we have
$$
\gamma^+_1+\gamma^+_2+\gamma^-_1+\gamma^-_2
=\left.\left[\frac{\partial\overline{\theta}}{\partial x}\right]
\right|_{x=\hat{\varphi}_1(t)}
+\left.\left[\frac{\partial\overline{\theta}}{\partial x}\right]
\right|_{x=\hat{\varphi}_2(t)}
=2\psi'_{0t}.
$$

We transform the left-hand side of relation
(02):
\begin{align*}
I&\buildrel{\rm def}\over=\psi_{0t}\left[\rho_{\tau}
-\frac{1}{2}\int_{\mathbb{R}^{1}}
\dot{\omega}_{0}(z)\omega_{0}(-z-\eta)
\left(\rho_{\tau}-\frac{2\beta_{\tau}}{\beta^{2}}z\right)dz\right]\\
&=\psi_{0t}\left[\rho_{\tau}+
\frac{1}{2}\int_{\mathbb{R}^{1}}
\dot{\omega}_{0}(z)\omega_{0}(-z-\eta)
\frac{\beta_{\tau}}{\beta^{2}}z\,dz\right.\\
&\left.
-\frac{1}{2}\int_{\mathbb{R}^{1}}
\dot{\omega}_{0}(z)\omega_{0}(-z-\eta)
\left(\rho_{\tau}+\frac{\beta_{\tau}}{\beta^{2}}(-z-\eta)+
\frac{\beta_{\tau}}{\beta^{2}}\eta\right)\right].
\end{align*}
Next, we change the variables in the last integral
\begin{equation*}
\frac{\beta_{\tau}}{\beta^{2}}
\int_{\mathbb{R}^{1}}(-z-\eta)\dot{\omega}_{0}(z)
\omega_{0}(-z-\eta)\,dz
=-\frac{\beta_{\tau}}{\beta^{2}}
\int_{\mathbb{R}^{1}}
z\dot{\omega}_{0}(-z-\eta)\omega_{0}(z)\,dz
\end{equation*}
and note that
\begin{equation*}
\rho_{\tau}-\frac{\beta_{\tau}}{\beta^{2}}\eta
=\frac{1}{\beta}\frac{\partial}{\partial \tau}\eta.
\end{equation*}
Finally, we get
\begin{align*}
I&=\psi_{0t}\left[\rho_{\tau}-
\frac{1}{2}\int_{\mathbb{R}^{1}}
\left\{z\dot{\omega}_{0}(z)\omega_{0}(-z-\eta)
+z\dot{\omega}_{0}(-z-\eta)\omega_{0}(z)\right\}\,dz\right.\\
&\qquad -\left.\frac{1}{2\beta}\frac{\partial}{\partial
\eta}\int_{\mathbb{R}^{1}}
\dot{\omega}_{0}(z)\omega_{0}(-z-\eta)\,dz\right].
\end{align*}
Since the function $\omega_{0}(z)$ is odd, we see that
the expression in braces
in the first integral in the right-hand side  is
\begin{equation*}
-z\frac{\partial}{\partial z}
\left(1-\omega_{0}(z)\omega_{0}(z+\eta)\right).
\end{equation*}
Hence, integrating by parts, we obtain
\begin{align*}
I&=\psi_{0t}\left[\rho_{\tau}-\frac{\beta_{\tau}}{2\beta^{2}}
\int_{\mathbb{R}^{1}}
\left(1-\omega_{0}(z)\omega_{0}(z+\eta)\right)\,dz\right.\\
&\qquad -\left.\frac{1}{2\beta}\frac{\partial\eta}{\partial\tau}
\int_{\mathbb{R}^{1}}
\dot\omega_{0}(z+\eta)\omega_{0}(z)\,dz\right].
\end{align*}
Or, finally,
\begin{equation}\label{I:1}
I=\psi_{0t}\left[\rho_{\tau}
+2\frac{1}{\beta}\frac{\partial}{\partial\tau}\tilde{B}\right],
\end{equation}
where
\begin{equation*}
\tilde{B}(\eta)
=\int_{\mathbb{R}^{1}}
\left(1-\omega_{0}(z+\eta)\omega_{0}(z)\right)\,dz
=2\eta\tanh\eta.
\end{equation*}
So, in view of (46), we obtain the equation for $\rho$:
\begin{equation}\label{RhoSt:1}
\frac{\partial}{\partial\tau}\left(\rho
+\frac{1}{2\beta}\tilde{B}(\eta)\right)=(\tau).
\end{equation}
Or
\begin{equation*}
\frac{\partial}{\partial\tau}
\left(\beta^{-1}(\eta+\eta\tanh\eta)\right)=2B.
\end{equation*}
After integration, we obtain
\begin{equation}
\eta(1+\tanh\eta)=
2\beta\int^{\tau}_{-\infty}B(\tau')\,d\tau'.
\end{equation}

Since $B(\tau)\to1$ as $\tau\to\infty$, 
we have $\eta\tau^{-1}\to1$ as $\tau\to\infty$.

Since $B(\tau)=\mathcal{O}(|\tau|^{-N})$ for any~$N$ 
as $\tau\to-\infty$, 
the integral in the left-hand of (49) converges 
and hence the right-hand side tends to the limit
$$
\lim_{\tau\to-\infty}\eta(1+\tanh\eta)
=\lim_{\tau\to-\infty}2\beta\int^{\tau}_{-\infty}B(\tau')\,d\tau'=0.
$$
Moreover, in view of the inequality $B\geq0$, 
we have $\eta\geq0$ for $\tau\in\mathbb{R}^2_\tau$,
which implies that 
$$
\eta\to0\qquad\text{as}\quad \tau\to-\infty.
$$
Here we took into account the inequality $\beta>0$, 
which follows from~(43).

Substituting $\beta=(\hat{C}\hat{D}^{-1}(\eta))^{1/2}$ into
(49), we obtain the following equation for the function $\eta$:
\begin{equation}
\eta(1+\tanh\eta)
=(\hat{C}\hat{D}^{-1})^{1/2}\int^{\tau}_{0}B\,d\tau'.
\end{equation}

Since the functions contained in (50) are monotone and the
limits exist as $\tau\to\pm\infty$, this equation is obviously
solvable.

Next, we have
$$
\beta(\tau)=\sqrt{\frac{\hat{C}(\eta)}{\hat{D}(\eta)}}\to\beta^-
=\mathrm{const},\qquad \tau\to-\infty.
$$
Moreover, as readily follows from the exponential rate
of convergence to the limit of the functions
$\hat{C}/\hat{D}$, $1-B$, and $1+\tanh\eta$ as
$\tau\to\pm\infty$, the derivatives satisfy the estimate
\begin{equation}
\frac{\partial^\alpha\beta}{\partial\tau^\alpha}
=\mathcal{O}(|\tau|^{-N}),\qquad |\tau|\to\infty.
\end{equation}
Now we calculate the limits of the expressions
$$
J_i=B(\gamma^+_i+\gamma^-_i)-A_i,\qquad i=1,2,
$$
as $\tau\to\pm\infty$.
As $\tau\to\infty$, we have
$$
\rho\to\infty\quad(\rho\sim\tau),\qquad B\to1,\qquad
A_i\to(-1)^{i+1}\varphi_{i0t}.
$$
Therefore, we have
$$
\lim_{\tau\to\infty}J_i=0
$$
in view of the Stefan conditions (9),
and $J_i=\mathcal{O}(\tau^{-1})$, $i=1,2$.

As $\tau\to-\infty$, we proved that $\eta\to0$
and hence 
\begin{equation}
\varphi_1-\varphi_2=\mathcal{O}(\varepsilon),\qquad 
\tau\to-\infty.
\end{equation}

\begin{theorem}
Relation {\rm(50)} is a sufficient condition for relation
{\rm(28)} to hold.
\end{theorem}

\begin{proof} In view of the corollary of Lemma~2, Section~5,
the estimate for the coefficients of the $\delta$-functions 
in~(28) as $\tau\to\infty$ and (52) 
are sufficient conditions for relation (28) to follow from
(46).  In turn, relation (50) follows from (46).
\end{proof}

Now we analyze the relations $V^2_i=0$ and (44).
We use the explicit form of the function $F(z)$ and the formula
\begin{equation}
\omega_0=\frac{e^z-e^{-z}}{e^z+e^{-z}}.
\end{equation}

As $\eta\to\infty$, we have
\begin{equation}
\Omega(z,\eta)=\omega_0(z)+\mathrm{O}(e^{-2\eta})
\end{equation}
and hence $\beta\to1$ as $\eta\to\infty$.

From relation (33) we easily obtain
\begin{gather*}
\Omega'_z(z,\eta)=\mathcal{O}(e^{-|z|}),\qquad |z|\to\infty,\\
\Omega'_\eta(z,\eta)=\mathcal{O}(e^{-|z|}),\qquad |z|\to\infty,\\
C_\Omega=\int_{\mathbb{R}^{1}} \dot\omega_0(z)(1-\omega_0(-z-\eta))\,dz\geq0.
\end{gather*}

Similarly to (54), using (45), we can verify the relations
\begin{gather}
\beta_\tau=\mathcal{O}(e^{-2|\eta|}), \qquad  \eta\to\infty,
\notag\\
B_\Omega=
1+\mathcal{O}(e^{-2|\eta|}),\quad\quad\eta\to\infty,
\\
C_\Omega=4+\mathcal{O}(e^{-2\eta}),\qquad \eta\to\infty,
\notag\\
B^z_\Omega=0,\qquad \eta\to\infty.
\notag
\end{gather}

We rewrite the expressions for the coefficients $V^1_i$ in more detail:
\begin{align*}
V^1_1&=-\varphi_{1t}(B_\Omega+C_\Omega)
+\frac{\beta'_\tau\psi'_{0t}}{\beta^2}B^z_\Omega-q\big|_{x=\varphi_1} C_\Omega,
\\
V^1_2&=-\varphi_{2t}(B_\Omega+C_\Omega)
+\frac{\beta'_\tau\psi'_{0t}}{\beta^2}B^z_\Omega+q\big|_{x=\varphi_2} C_\Omega.
\end{align*}
Form formulas (57), the definition of the functions $\varphi_i$,
and the kinetic overcooling conditions (9),
we obtain
$$
\lim_{\tau\to\infty} V^i_1=0,
$$
because $\check{q}|_{x=\varphi_i}=0$, while
$\hat{q}\equiv0$ for $t<t^*$, i.e.,
because of the fact that the model function $\check{T}+eI$ is
constructed so that conditions~(8) be satisfied.
Thus, in view of the corollary of Lemma~2, Section~5, for the relation
$$
V^1_1\delta(x-\varphi_1)+V^2_1\delta(x-\varphi)
=\mathcal{O}_{\mathcal{D}'}(\varepsilon)
$$
to hold, it is sufficient to have $V^1_1+V^2_2=0$ or,
in more detail,
\begin{equation}
(\varphi_{1t}+\varphi_{2t})\left(B_\Omega+C_\Omega\right)
-\frac{\beta_\tau \psi'_0}{\beta^2}B^z_\Omega
+(q|_{x=\varphi_1}-q|_{x=\varphi_2})C_\Omega=0.
\end{equation}
Thus, we have proved the following assertion.

\begin{theorem}
The conditions of Theorem~{\rm1} and relations {\rm(49), (55),
(56)} are sufficient for the functions $\check{\theta}$,
$\check{u}$ to be a weak asymptotic solution of the phase field
system.
\end{theorem}

We consider relation (56).
It follows from the above that
$C_\Omega=C_\Omega(\eta)$ decreases sufficiently fast
as $\eta\to\infty$, in any case
$|\,|\eta| C_\Omega(\eta)|\leq\mathrm{const}$.
This fact, the estimate
$$
\hat{q}|_{x=\varphi_1}-\hat{q}|_{x=\varphi_2}
=\mathcal{O}(|\varphi_1-\varphi_2|^\mu),\qquad
\mu\in(0,1/2),\quad t\leq t^*.
$$
proved in Lemma~4 in Section~5, 
relation (52) implies that the obvious estimate
$$
||\varphi_1-\varphi_2|^\mu C_\Omega|
\leq
\varepsilon\mathrm{const}|\eta C_\Omega|
=\mathcal{O}(\varepsilon^\mu)
$$
for $t\in[0,t_1]$.

For $t\leq t^*$, by Lemma~6, we have
$$
q\big|_{x=\varphi_i}-q\big|_{x=\varphi_{i0}}
=\mathcal{O}(|\psi_0\psi_1|^\mu),
$$
where $\psi_1=\varphi_{21}-\varphi_{11}$. Hence, by (15),
and a statement similar to Lemma~3 in Section~5 
(see also (17)), we have
$$
\big|2(\varphi_{10t}+\varphi_{20t})-q|_{x=\varphi_1}+q|_{x=\varphi_2}\big|
=\mathcal{O}(\varepsilon^\mu),\qquad t\leq t^*.
$$
We introduce a function $V(\tau)\in C^\infty$ such that
$V'_\tau\in S(\mathbb{R}^1)$, $V(-\infty)=0$, $V(\infty)=1$.
Then, in view of considerations similar to those used in Lemma~3,
we can show that the following estimate holds:
$$
(B_\Omega+C_\Omega)( V(\tau)(\varphi_{10t}+\varphi_{10t})
+(q|_{x=\varphi_2}-q|_{x=\varphi_1}) C_\Omega
=\mathcal{O}(\varepsilon^\mu),\quad \mu\in(0,1/2),
$$
for $t\leq t^*$.

This implies that the left-hand side of (32) is estimated as
$\mathcal{O}_{\mathcal{D}'}(\varepsilon^\mu)$ if
\begin{equation}
(B_\Omega+C_\Omega)[-V(\varphi_{10t}+\varphi_{20t})
+\varphi_{1t}+\varphi_{2t}]
-2\frac{\beta'_\tau\psi'_{0t}}{\beta^2}B^z_\Omega=0.
\end{equation}
From this relation we obtain
\begin{equation}
\frac{\partial}{\partial\tau}(\tau(\varphi_{11}+\varphi_{21}))
=(V-1)(\varphi_{10t}+\varphi_{20t})
+2\frac{\beta'_\tau\psi'_{0t}}{\beta^2}
\frac{B^z_\Omega}{B_\Omega+C_\Omega}.
\end{equation}
From this equation we determine the extensions
of the functions $\varphi_{11}$ and $\varphi_{21}$.
We note that our argument results in an equation
that does not contain the temperature, 
namely,
the assumption that the classical solution exists till
$t=t^*$ is sufficient for constructing 
a global smooth approximation of the solution. 

Let us calculate the function $B_\Omega=B_\Omega(\eta)$ 
in more detail. We have
$$
B_\Omega(\eta)=\int\Omega'_z(\Omega'_z-\Omega'_\eta)\,dz.
$$
In this integral, we make the change of variable
$z\to-z-\eta$. Then we obtain
$$
\Omega'_\eta\to -(\Omega'_z-\Omega'_\eta), \qquad
(\Omega'_z-\Omega'_\eta)\to-\Omega'_z.
$$
Hence we have 
$$
B_\Omega(\eta)=\int \Omega'_z \Omega'_\eta\,dz.
$$
Finally, we obtain 
$$
B_\Omega=\frac12\int\big(\Omega'_z(\Omega'_z-\Omega'_\eta)
+\Omega'_z\Omega'_\eta\big)\,dz
=\frac12\int(\Omega'_z)^2\,dz.
$$
Thus, for finite $\eta$, the denominator
$(B_\Omega+C_\Omega)$ in the last term 
in the right-hand side of~(58) does not vanish.
Moreover, using the explicit form of the functions
$\omega_0(z)$, $B^z_\Omega$, $B_\Omega$, and $C_\Omega$, 
we can verify that  
$$
2\frac{\beta'_\tau B^z_\Omega}{\beta^2}
=\mathcal{O}(|\eta|^{-N}),\qquad |\tau|\to\infty,
$$
where $N\gg1$ is an arbitrary number.

Therefore, we have
\begin{equation}
\varphi_{11}+\varphi_{12}=
\frac{\varphi_{10t}+\varphi_{20t}}{\tau}
\int^{\tau}_{0}(V(\tau')-1)\,d\tau'
+\frac{2\psi'_{0t}}{\tau}\int^{\tau}_{0}
\frac{\beta'_\tau}{\beta^2}
\frac{B^z_\Omega}{B^z_\Omega+C_\Omega}\,d\tau'.
\end{equation}
We note that 
\begin{equation}
\psi'_{0t}\bigg(1+\frac{\partial}{\partial\tau}
[\tau(\varphi_{21}-\varphi_{11})]\bigg)=\psi'_0\rho_\tau,
\end{equation}
where $\rho=\eta\beta^{-1}$ and the functions $\eta$ and $\beta$
are determined by Eqs.~(50) and~(43).

The system of Eqs.~(61) and (62) allows one to find the
functions $\varphi_{i1}$, $i=1,2$, and thus completely
determines the functions contained in the ansatz
of the weak asymptotic solution of system~(1).

It is easy to see that the solutions thus constructed
satisfy conditions~(15).

Next, using the explicit formulas for
$B^z_\Omega$, $B_\Omega$, and $C_\Omega$,
we can easily verify that $B^z_\Omega|_{\eta=0}=0$
(because $\omega_0(z)$ is odd) and
$(B_\Omega+C_\Omega)|_{\eta=0}\not=0$,
because $\dot\omega_0(z)$ is even.
This means that
\begin{equation}
(\varphi_{1t}+\varphi_{2t})|_{\eta=0}=0.
\end{equation}

We note that the right-hand side of (27) is a piecewise
smooth function for $|\psi|\geq\mathrm{const}>0$
and smooth for $x\not=\varphi_i$, $i=1,2$.
As $\psi\to0$ ($t\to t^*$), the right-hand side of (27)
becomes proportional to $\delta(x-x^*)$, where
$x^*_0=x^*|_{\psi=0}$,
and the proportionality coefficient is negative.
Thus, if $x$ and $t$ vary in the respective neighborhoods
of the points $x^*_0$ and $t^*$, then the $\delta$-function
with a negative coefficient appears and disappears in the
right-hand side of (27), which results in a negative
soliton-like jump of the temperature in a neighborhood
of the point $t=t^*$, $x=x^*_0$.

{\it Calculation of the temperature jump}.
To prove the statement in the Introduction 
concerning the temperature jump, 
we must calculate the quantity
$$
[eI+\hat{q}+\check{q}]|_{t=t^*,\,x=x^*}.
$$

Let us verify that $\check{q}+\hat{q}$ 
is a continuous function.
This function is the sum of solutions 
of the heat equation 
with singular right-hand sides,
but these singularities arise 
after the substitution of the continuous function
$e{\check{T}}$ (which, of course, is not a solution) 
into the left-hand side.
It remains to prove that the solution differs 
from the function $e{\check{T}}$ by a continuous function.  
For this, 
it suffices to verify that the right-hand sides,
which arise after the substitution of $e{\check{T}}$ 
into the equation, 
do not generate any singularities 
in the solution of the heat equation
in addition to those contained 
in the function~$e{\check{T}}$.

Here we, in contrast to the preceding statements, 
use the fact that the singularities 
in the right-hand sides 
arise as the result of the substitution. 

By $\check{q}_i$, $i=1,2$, we denote the terms 
in $\hat{q}+ \check{q}$ corresponding to the right-hand sides:  
\begin{align*}
f_1&=-\bigg(\frac{\partial^2}{\partial x^2}\gamma^-
\frac{(\varphi_1-x)(x-\varphi_2)}{\psi} \bigg)
B(\tau)[H(x-\varphi_1)-H(x-\varphi_2)],
\\
f_2&=-\bigg(\frac{\partial^2}{\partial x^2}\hat{\gamma}
\frac{(\varphi_1-x)(x-\varphi_2)}{\psi} \bigg)
(1-B(\tau))[H(x-\varphi_2)-H(x-\varphi_1)],
\end{align*}
The other terms in $q$ are solutions of the heat 
equation with piecewise continuous right-hand side and hence
are continuous.

The functions $\hat{q}_i$ have a similar property if 
they are calculated up to $\mathcal{O}(\varepsilon)$. 
Indeed, we denote 
$$
\Pi=\gamma^-\frac{(\varphi_1-x)(x-\varphi_2)}{\psi}
$$
and represent, for example, 
the function $\hat{q}_1$ in the form
$$
\hat{q}_1=-\frac1{2\sqrt{2\pi}}
\int^t_0\frac{B(\tau(t',\varepsilon))}{\sqrt{t-t'}}
\int^{\varphi_2}_{\varphi_1}\frac{\partial^2\Pi}{\partial \xi^2}
e^{-\frac{(x-\xi)^2}{4(t-t')} }\,d\xi dt'.
$$

Omitting the number factor and integrating by parts
in the integral over~$\xi$, we obtain 
\begin{align*}
\hat{q}_1&=-\int^t_0\frac{B}{\sqrt{t-t'}}
e^{-\frac{(x-\xi)^2}{4(t-t')}}
\bigg|^{\varphi_2}_{\varphi_1}
-\int^t_0\frac{B}{\sqrt{t-t'}}\frac{\partial^2}{\partial \xi^2}
e^{-\frac{(x-\xi)^2}{4(t-t')}}\,d\xi dt'
\\
&=-\int^t_0\frac{B}{\sqrt{t-t'}}
\Big[e^{-\frac{(x-\xi)^2}{4(t-t')}}\gamma^-_2
-e^{-\frac{(x-\xi)^2}{4(t-t')}}\gamma^+_1\Big]\,dt'
\\
&\qquad 
+\int^t_0\frac{\partial}{\partial t'}
\bigg(\int^{\varphi_2}_{\varphi_1}
\frac{\Pi e^{-\frac{(x-\xi)^2}{4(t-t')}}}{\sqrt{t-t'}}\,d\xi\bigg)\,dt'
+\int^t_0\bigg(\int^{\varphi_2}_{\varphi_1} 
B \frac{\partial \Pi}{\partial t'}
\frac{e^{-\frac{(x-\xi)^2}{4(t-t')}}}{\sqrt{t-t'}}\,d\xi\bigg)\,dt'.
\end{align*}

The last term is a solution of the heat equation 
with the piecewise continuous right-hand side
$$
B\frac{\partial \Pi}{\partial t}
[H(x-\varphi_1)-H(x-\varphi_2)], 
$$
and hence it is continuous 
uniformly w.r.t $\varepsilon\geq0$. 
The other terms are also continous functions
uniformly w.r.t $\varepsilon\geq0$. 

It is easy to see that in the formula for $\hat{q}_1$
there is no term containing the derivative 
$\frac{\partial B}{\partial t}$, 
because this term would be of order $\mathcal{O}(\varepsilon)$.

Indeed
$$
\bigg|\frac{\partial B}{\partial t}\Pi
[H(x-\varphi_1)-H(x-\varphi_2)]\bigg|
\leq C|B'\cdot\psi'_{0t}\cdot\rho|,
$$
since $|\Pi|\leq C\psi$ for $x\in[\varphi_1,\varphi_2]$ 
and $\varepsilon^{-1}\psi=\rho$.
The derivative $B'_\rho$ decreases faster than any power 
of $|\rho|^{-1}$, $|\rho'_\tau|\leq\mathrm{const}$. 
This implies that 
$$
\int^t_0\frac{\partial B}{\partial t}\,dt'
=\mathcal{O}(\varepsilon). 
$$

Hence we have
$$
[\check{q}+\hat{q}]|_{x=x^*,\,t=t^*}=0.
$$
Let us calculate the function $eI$.
We have
$$
eI|_{x=x^*}=\frac12(\varphi_{1t}-\varphi_{2t})
=\frac{\psi'_0}{2}\dot\rho_\tau.
$$
It follows from (48) that 
\begin{gather*}
\dot\rho_\tau\to1,\qquad \tau\to\infty,
\\
\dot\rho_\tau\to0,\qquad \tau\to-\infty
\end{gather*}
and hence 
$$
[eI]|_{\substack{x=x^*,\\t=t^*}}
=-\frac12\lim_{t\to t^*-0}(\varphi_{10t}+\varphi_{20t}).
$$

\section{Technique of the weak asymptotics\\ method}\label{WAsMeth}

First, we recall the definition of
\emph{regularization of the generalized function}.

\begin{definition}\label{Def:Regul}
A family of functions $f(x,\varepsilon)$
smooth for $\varepsilon>0$ and satisfying the condition
\begin{equation*}
\underset{\varepsilon\to0}{\mathrm{w}-\lim}f(x,\varepsilon)=f(x)
\end{equation*}
is called the regularization of the generalized function$f(x)$.
\end{definition}

We note that, by definition,
the last relation can be rewritten as
\begin{equation*}
\lim_{\varepsilon\to0}\langle
f(x,\varepsilon),\zeta(x)\rangle=\langle f,\zeta\rangle
\end{equation*}
for any test function $\zeta(x)$
(from now on, $\langle\ ,\ \rangle$
denotes the action of a generalized function
on a test function).

\begin{lemma}
Let $\Gamma_{t}=\{x-\varphi(t)=0\}$,
$x\in\mathbb{R}$, where $\varphi(t)$ is a smooth function,
let $\omega(z)\in\mathbb{S}$
($\mathbb{S}$ is the Schwartz space),
and let $\beta=\beta(t)>0$.
Then the following relation holds for any test function
$\zeta(x)${\rm:}
$$
\frac{1}{\varepsilon}
\Big\langle\omega\left(\beta\frac{x-\varphi(t)}{\varepsilon}\right),
\zeta(x)\Big\rangle
=\frac{1}{\beta}A_{\omega}\zeta(\varphi)+\mathcal{O}(\varepsilon),
$$
where $A_{\omega}=\int^{\infty}_{-\infty}\omega(z)dz$.
\end{lemma}

\begin{proof}
The expression in the right-hand side can be written as
\begin{equation*}
\frac{1}{\varepsilon}
\int_{\mathbb{R}^{1}}\omega
\left(\beta\frac{x-\varphi(t)}{\varepsilon}\right)\zeta(x)\,dx
= \frac{\zeta(\varphi)}{\beta}\int_{\mathbb{R}^{1}}\omega(z)\,dz
+\mathcal{O}(\varepsilon).
\end{equation*}
Here we perform the change of variables
$z=\beta(x-\varphi)/\varepsilon$ and apply the Taylor formula
to the integrand at the point $x=\varphi$.
By definition, the last integral is the action of the
generalized function
$\beta^{-1}A_{\omega}\delta(x-\varphi)$ on the test function $\zeta$.
\end{proof}

\textbf{Proof of the formula in Example~1 in the
Introduction.}
Let $\omega(z) \in C^{\infty}$,
$\omega'\in\mathbb{S}(\mathbb{R}^{1})$,
$\lim_{z\to+\infty}\omega(z)=0$,
and
$\lim_{z\to-\infty}\omega(z)=1$.
We verify that
\begin{equation}\label{HW:1}
\omega\left(\frac{x-x_{0}}{\varepsilon}\right)-H(x-x_{0})
=\mathcal{O}_{\mathcal{D}'}(\varepsilon),
\quad x_{0}=\mathrm{const},
\end{equation}
where $H$ is the Heaviside function.
By Definition~\ref{WkMDef},
we consider the expression
\begin{align}\label{HW:2}
&\int_{\mathbb{R}^{1}}
\left[\omega\left(\frac{x-x_{0}}{\varepsilon}\right)
-H(x-x_{0})\right]\zeta(x)\,dx\\
&\qquad
=\varepsilon\zeta(x_{0})
\int_{\mathbb{R}^{1}}\left[\omega(z)-H(z)\right]\,dz
+\mathcal{O}(\varepsilon^{2}).\notag
\end{align}
Here we performed the change of variables
$z=(x-x_{0})/\varepsilon$ and applied the Taylor formula
to the functions $\zeta(x)$
at the point $x=x_{0}$.
In view of our assumptions and the properties of the
function~$H$, the integral in the right-hand side of the last
relation converges and hence the right-hand side of (\ref{HW:2})
is of order  $\mathcal{O}(\varepsilon)$.
We thus obtain estimate (\ref{HW:1}).
\medskip

\textbf{Proof of the formula in Example~2 in the
Introduction.}
Let $\omega_{i}(z) \in C^{\infty}$,
$\left(\omega_{i}\right)'\in\mathbb{S}(\mathbb{R}^{1})$,
$\lim_{z\to+\infty}\omega_{i}(z)=0$,
and $\lim_{z\to-\infty}\omega_{i}(z)=1$, $i=1,2$.
We consider the integral
\begin{align*}
J&\buildrel{\rm def}\over=
\int_{\mathbb{R}^{1}}
\omega_{1}\left(\frac{x-x_{1}}{\varepsilon}\right)
\omega_{2}\left(\frac{x-x_{2}}{\varepsilon}\right)\zeta(x)\,dx\\
&=\int_{\mathbb{R}^{1}}
\omega_{1}\left(\frac{x-x_{1}}{\varepsilon}\right)
\omega_{2}\left(\frac{x-x_{2}}{\varepsilon}\right)
\left(\int_{-\infty}^{x}\zeta(y)dy\right)'_x\,dx.
\end{align*}
Integrating by parts, we obtain
\begin{align*}
J&=-\int_{\mathbb{R}^{1}}
\dot{\omega}_{1}\left(\frac{x-x_{1}}{\varepsilon}\right)
\omega_{2}\left(\frac{x-x_{2}}{\varepsilon}\right)
\left(\int_{-\infty}^{x}\zeta(y)dy\right)\,dx\\
&\qquad
-\int_{\mathbb{R}^{1}}
\dot{\omega}_{2}\left(\frac{x-x_{2}}{\varepsilon}\right)
\omega_{1}\left(\frac{x-x_{1}}{\varepsilon}\right)
\left(\int_{-\infty}^{x}\zeta(y)dy\right)\,dx.
\end{align*}
We perform the change of variables $z=(x-x_{i})/\varepsilon$
with $i=1$ in the first integral
and with $i=2$ in the second integral
and apply the Taylor formula to the function
$F(x)=\int^{x}_{-\infty}\zeta(y)\,dy$
at the points $x=x_{i}$, $i=1,2$, respectively.
Then we calculate the first and the second integral,
we have
\begin{gather*}
J=B_{1}\left(\frac{\triangle
x}{\varepsilon}\right)H(x-x_{1})+B_{2}\left(\frac{\triangle x}{\varepsilon}\right)
H(x-x_{2})+\mathcal{O}_{\mathcal{D}'}(\varepsilon),\quad
\triangle x=x_{1}-x_{2},
\\
B_{1}(\rho)=\int_{\mathbb{R}^{1}}
\dot{\omega}_{1}(z)\omega_{1}(-z-\rho)\,dz,\quad
B_{2}(\rho)=\int_{\mathbb{R}^{1}}
\dot{\omega}_{2}(z)\omega_{1}(z-\rho)\,dz.
\end{gather*}

To calculate the linear combination of generalized functions
up to $\mathcal{O}_{\mathcal{D}'}(\varepsilon^{\alpha})$,
we must improve the classical definition of linear independence.
This improvement plays the key role in the study
of problems with interaction of nonlinear waves.

Indeed, let $\phi_{1}\neq\phi_{2}$ be independent of~$x$.
We consider the expression
\begin{equation}\label{LinDelta:1}
g_{1}\delta(x-\phi_{1})+g_{2}\delta(x-\phi_{2})
=\mathcal{O}_{\mathcal{D}'}(\varepsilon^{\alpha}),\qquad
\alpha>0,
\end{equation}
where the functions $g_{i}$ are independent of $\varepsilon$.
Clearly, for the last relation to be satisfied,
it suffices to have
\begin{equation*}
g_{i}=\mathcal{O}(\varepsilon^{\alpha}),\ \ \ i=1,2,
\end{equation*}
or, with the properties of the functions $g_{i}$ taken into account,
\begin{equation*}
g_{i}=0,\qquad i=1,2.
\end{equation*}

But, if we assume that the coefficients $g_{i}$
depend on the parameter $\varepsilon$,
then the above estimates do not work.
Namely, let us consider the following specific case
of this dependence:
\begin{equation}\label{g:g}
g_{i}=A_{i}+S_{i}(\triangle\phi/\varepsilon),\quad i=1,2,
\end{equation}
where $A_{i}$ are independent of $\varepsilon$,
the functions $S_{i}(\sigma)$ decrease sufficiently fast
as $|\sigma|\to\infty$, and $\triangle\phi=\phi_{2}-\phi_{1}$.

\begin{lemma}[\rm Linear independence
of generalized functions]\label{Lemma:2}
Suppose that the estimate
\begin{equation*}
|\sigma S_{i}(\sigma)|\leqslant \mathrm{const},\quad i=1,2, \quad
-\infty<\sigma<+\infty
\end{equation*}
holds. Then, for $\alpha=1$, expression {\rm(\ref{LinDelta:1})}
implies the relations
\begin{equation}\label{LinCoef:1}
A_{1}=0,\quad A_{2}=0,\quad S_{1}+S_{2}=0.
\end{equation}
\end{lemma}

\begin{proof}
Using the Taylor formula in (\ref{LinDelta:1})
and taking (\ref{g:g}) into account, we obtain
\begin{equation*}
S_{1}\zeta(\phi_{1})+S_{2}\zeta(\phi_{2})
=S_{1}\zeta(\phi_{1})+S_{2}\zeta(\phi_{1})
+S_{2}(\phi_{2}-\phi_{1})\zeta'(\phi_{1}+\mu\phi_{2}),
\end{equation*}
where $0<\mu<1$
Since the function $\sigma S_{2}(\sigma)$ is uniformly bounded
in $\sigma\in\mathbb{R}^{1}$, we obtain
\begin{equation*}
S_{2}(\triangle\phi/\varepsilon)(\phi_{2}-\phi_{1})
=\left.\{-\sigma S_{2}(\sigma)\}
\right|_{\sigma=\triangle\phi/\varepsilon}\cdot\varepsilon
=\mathcal{O}(\varepsilon).
\end{equation*}
So expression (\ref{LinDelta:1}) can be rewritten as
\begin{equation*}
A_{1}\zeta(\phi_{1})+A_{2}\zeta(\phi_{2})+(S_{1}+S_{2})\zeta(\phi_{1})
=\mathcal{O}(\varepsilon).
\end{equation*}
Thus, because the coefficients $A_{i}$
are independent of $\varepsilon$,
we obtain the assertion of  Lemma~\ref{Lemma:2}.
\end{proof}

\begin{corollary}
Suppose that 
$$
|\sigma S_i(\sigma)|\leq\mathrm{const},\qquad 
i=1,2,\quad\sigma\geq0
$$
and the functions $\phi_i=\phi_i(t,\varepsilon)$ 
are continuous and uniformly continuous and satisfy 
the condition that 
$\Delta\phi>0$ for $t>0$ and $\varepsilon\geq0$ 
and $\Delta\phi=\mathcal{O}(\varepsilon)$ for $t<0$.
Then, for $\alpha=1$, expression {\rm(66)} 
implies~{\rm(64)}.
\end{corollary}

\begin{proof} 
Following the above argument, we obtain 
$$
S_1\zeta(\phi_1)+S_2\zeta(\phi_2)
=(S_1+S_2)\zeta(\phi_1)
+S_2(\phi_2-\phi_1)\zeta'(\phi_1+\mu\phi_2).
$$

In view of our assumptions,
$$
S_2(\phi_2-\phi_1)=\mathcal{O}(\varepsilon)
$$
uniformly in~$t$, which implies the desired assertion.
\end{proof}

\begin{lemma}\label{Lemma:3}
Suppose that $f(t)\in\mathbb{C}^{1}$, $f(t_{0})=0$,
and $f'(t_{0})\neq 0$.
Suppose also that $g(t,\tau)$ locally uniformly in $t$
satisfies the condition
\begin{equation*}
|\tau g(t,\tau)|\leqslant \mathrm{const},\quad
|\tau g'_t(t,\tau)|\leqslant \mathrm{const},\quad -\infty<\tau<\infty,
\end{equation*}
and $g(t_{0},\tau)=0$. Then the inequality
\begin{equation*}
\left|g\left(t,f(t)/\varepsilon\right)\right|\leqslant \varepsilon
C_{\hat{t}},
\end{equation*}
where $C_{\hat{t}}=\mathrm{const}$,
holds in any interval $0\leqslant t\leqslant \hat{t}$ that does
not contain zeros of the function $f(t)$ except for~$t_{0}$.
\end{lemma}

\begin{proof}
The fraction $\frac{f(t)}{t-t_{0}}$ is locally bounded in~$t$.
The fraction $\frac{\tau g(t,\tau)}{t-t_{0}}$ 
is also locally bounded.
We have
\begin{equation*}
g\left(t,f(t)/\varepsilon\right)=
\varepsilon\cdot\frac{g\left(t,f(t)/\varepsilon\right)}{(t-t_{0})}\cdot
\frac{f(t)}{\varepsilon}\cdot\frac{t-t_{0}}{f(t)}.
\end{equation*}
According to the assumptions of the lemma,
the last factor in the right-hand side is bounded
on the interval under study.
The product of the first and second factors
(without $\varepsilon$) is bounded in view of the properties
of the function $g(t,\tau)$.
\end{proof}

\begin{corollary}\label{Corollary:2}
Suppose that the assumptions of Lemma~{\rm\ref{Lemma:3}}
are satisfied for $0\leqslant\tau<\infty$
$(-\infty<\tau\leqslant0)$.
Then the assertion of Lemma~{\rm\ref{Lemma:3}}
holds on any half-interval $(t_{0},\hat{t}]$,
which does not contain zeros of the function~$f(t)$,
and $\mathrm{sign} \hat{t}=\mathrm{sign} f(t)$,
$t\in (t_{0},\hat{t}]$.
\end{corollary}

The proof of Corollary~\ref{Corollary:2} is obvious.

\begin{lemma} 
The following inequality holds{\rm:}
$$
\hat{q}|_{x=\varphi_1}-\hat{q}|_{x=\varphi_2}
=\mathcal{O}(|\varphi_1-\varphi_2|^\mu).
$$
\end{lemma}

\begin{proof}
We consider only one of the the functions $\hat{q}$ from (29),
namely, the functions determined by relation (30),
and apply the method developed in \cite{14}.
We consider teh difference of expressions (30)
omitting the number factors:
\begin{align*}
&\hat{q}|_{x=\varphi_1}-\hat{q}|_{x=\varphi_2}
\\
&\qquad=\int^t_0 \int^{\varphi_2}_{\varphi_1}
\frac{\displaystyle
[\exp\frac{(\varphi_1-\xi)^2}{4(t-t')}
-\exp(-\frac{(\varphi_2-\xi)^2}{4(t-t')})]}
{\psi\sqrt{t-\tau}}\,d\xi\,d\tau
\\
&\qquad=\int^t_0 \int^{\varphi_2}_{\frac{\varphi_1+\varphi_2}2}
\frac{\displaystyle
\exp(-\frac{(\varphi_2-\xi)^2}{4(t-t')})
[\exp\frac{(\varphi_2-\varphi_1)(\varphi_1+\varphi_2-2\xi)}{4(t-t')}
-1]} {\psi\sqrt{t-t'}}
\\
&\qquad\qquad\times
\frac{(t-t')^\mu}{(\varphi_2-\varphi_1)^\mu}
\frac{(\varphi_2-\varphi_1)^\mu}{(t-t')^\mu}
\,d\xi\,d\tau
\\
&\qquad=\int^t_0 \int_{\varphi_1}^{\frac{\varphi_1+\varphi_2}2}
\frac{\displaystyle
\exp(-\frac{(\varphi_1-\xi)^2}{4(t-t')})
[\exp\frac{(\varphi_1-\varphi_2)(\varphi_1+\varphi_2-\xi)}{4(t-t')}
-1]} {\psi\sqrt{t-t'}}
\\
&\qquad\qquad\times
\frac{(t-t')^\mu}{(\varphi_2-\varphi_1)^\mu}
\frac{(\varphi_2-\varphi_1)^\mu}{(t-t')^\mu}
\,d\xi\,dt',
\end{align*}
where we choose $\mu\in(0,1/2)$.

Next, using the inequality
$$
\bigg|\frac{e^{-\alpha x}-1}{x^\mu}\bigg|\leq\mathrm{const}
$$
for $\alpha>0$, $x\in[0,\infty)$,
and taking into account the fact that
the integral
$$
\int^t_0 (t-t')^{-(1/2+\mu)}\,dt',\qquad
\mu\in(0,1/2),
$$
converges, we obtain the statement of the lemma
for $\varphi_2\geq\varphi_1$. 
If $\varphi_2\leq\varphi_1$ (this is true for $t\geq t^*$), 
then we must change the exponents outside the square brackets 
in the integrands.
\end{proof}

\begin{lemma} 
Suppose that $\omega(z)\in C^\infty$
decreases faster than any power of $|z|^{-1}$ as $|z|\to\infty$.
Then
$$
\varepsilon^{-1}\omega((x-\varphi_i)/\varepsilon) \hat{q}_j
=\hat{q}_j\delta(x-\varphi_i)\int\omega(z)\,dz
+\mathcal{O}_{\mathcal{D}'}(\varepsilon^\mu),
$$
where $\hat{q}_j$ is one of the functions defined by the
relations {\rm(29)}, $i=1,2$, $\mu\in(0,1/2)$.
\end{lemma}

\begin{proof}
We consider one of the functions $\hat{q}$, namely,
the function defined by the relations
$$
\hat{q}=\frac1{2\sqrt{\pi}}\int^t_0\frac{dt'}{\psi\sqrt{t-t'}}
\int^{\varphi_2}_{\varphi_1}\exp(-\frac{(x-\xi)^2}{4(t-t')})\,d\xi.
$$
The desired relation can be written as
$$
\varepsilon^{-1}\int \zeta(x)
\omega\bigg(\frac{x-\varphi_i}{\varepsilon}\bigg)\hat{q}\,dx
=\zeta(\varphi_i)\hat{q}|_{x=\varphi_i}
\int\omega(z)\,dz
+\mathcal{O}(\varepsilon^\mu)
$$
or, omitting the number factors, in the form
\begin{align}
I&\od \varepsilon^{-1}\int\zeta(x)
\omega\bigg(\frac{(x-\varphi_i)}{\varepsilon}\bigg)\hat{q}\,dx
\\
&\qquad\times
\int^t_0 \frac{dt'}{\psi\sqrt{t-t'}}
\int^{\varphi_2}_{\varphi_1}
\exp\bigg(-\frac{(x-\xi)^2}{4(t-t')}\bigg)\,d\xi\,dx
\notag\\
&=\zeta(\varphi_i)\int^t_0 \frac{dt'}{\psi\sqrt{t-t'}}
\int^{\varphi_2}_{\varphi_1}
\exp\bigg(-\frac{(\varphi_i-\xi)^2}{4(t-t')}\bigg)\,d\xi
+\mathcal{O}(\varepsilon^\mu).
\notag
\end{align}
In the left-hand side, we change the variables
$(x-\varphi_i)/\varepsilon=z$ and obtain
$$
I=\int\zeta(\varphi_i+\varepsilon z)\omega(z)
\int^t_0 \frac{dt'}{\psi\sqrt{t-t'}}
\int^{\varphi_2}_{\varphi_1}
\exp\bigg(-\frac{(\varphi_i-\xi+\varepsilon z)^2}{4(t-t')}\bigg)\,
dz\,d\xi.
$$
It is clear that to prove relation (67),
it suffices to prove that, with accuracy up to small values,
the term $\varepsilon z$ can be omitted in the exponent.

We consider the expression
\begin{align*}
J&=\int\zeta(\varphi_i+\varepsilon z)\omega(z)
\int^t_0\frac{dt'}{\psi\sqrt{t-t'}}
\\
&\qquad\times
\int^{\varphi_2}_{\varphi_1}
\bigg[\exp\bigg(-\frac{(\varphi_i-\xi+\varepsilon z)^2}{4(t-t')}\bigg)
-\exp\bigg(-\frac{(\varphi_i-\xi)^2}{4(t-t')}\bigg)\bigg]\,dz\,d\xi.
\end{align*}
Since the function $\omega$ decreases fast,  we can assume
that $|z|<\varepsilon^{-\delta}$, $\delta\in(0,1)$.
Next it suffices to consider the integral over $t'$
from $0$ to $t-\varepsilon^{1-\delta}$, because if
$t'\in[t-\varepsilon^{1-\delta},t]$, then the obtained integral
can be estimated as $\mathcal{O}(\varepsilon^{(1-\delta)/2})$.

In the remaining integrals, we can use the same method as
in the proof of Lemma~4. Namely, we transform the difference
of the exponential functions:
\begin{align*}
&\exp\bigg(-\frac{(\varphi_i-\xi+\varepsilon z)^2}{4(t-t')}\bigg)
-\exp\bigg(-\frac{(\varphi_i-\xi)^2}{4(t-t')}\bigg)
\\
&\qquad
=\bigg[\exp\bigg(-\frac{(2(\varphi_i-\xi)-\varepsilon z)\varepsilon z}
{4(t-t')}\bigg)-1 \bigg]
\bigg(\frac{\varepsilon z}{t-t'}\bigg)^\gamma
\bigg(\frac{\varepsilon z}{t-t'}\bigg)^{-\gamma},
\\
&\qquad
\gamma\in(0,1/2),
\end{align*}
and apply the estimate
$$
\bigg|\bigg(\exp\bigg(
-\frac{(2(\varphi_i-\xi)-\varepsilon z)\varepsilon z}
{4(t-t')}\bigg)-1\bigg)\bigg/
\bigg(\frac{t-t'}{\varepsilon z}\bigg)^\gamma\bigg|<\mathrm{const},
$$
which, obviously, is valid in the required range of variables.
Hence, as in Lemma~4, we obtain the estimate
$$
J=\mathcal{O}(\varepsilon^\mu),\qquad \mu\in(0,1/2),
$$
which proves relation~(67).
\end{proof}

Similarly to Lemmas~4 and~5, we prove the following 
statement.

\begin{lemma}
The following relation holds for any of the funtions 
$\hat q_i$ defined by relations {\rm(29):}
$$
\hat q_i(\varphi_i,t)-\hat q_i(\hat q_{i0},t)
=\mathcal{O}(|\psi_0\psi_1|^\mu),\qquad \mu\in(0,1/2).
$$
\end{lemma}

\begin{lemma} 
Relations {\rm(38), (39)} hold.
\end{lemma}

\begin{proof}
In Section~3, we introduced the function
$\Omega(z,\eta)$.
Of course, we can change the signs in the arguments
using the fact that the function $\omega_0(z)$ is odd,
but the form presented above is convenient for calculations.

Calculating the derivatives $u_t$ and $u_x$, we obtain
\begin{align}\label{UdifT:1}
&\check{u}_{t}=\frac{1}{2\varepsilon}
\left\{\left[\beta(\varphi_{1}-x)\right]_{t}
\dot{\omega}_{0}\left(\beta\frac{\varphi_{1}-x}{\varepsilon}\right)
+\left[\beta(x-\varphi_{2})\right]_{t}
\dot{\omega}_{0}\left(\beta\frac{x-\varphi_{2}}{\varepsilon}\right)
\right.\notag\\
&\qquad-\left[\beta(\varphi_{1}-x)\right]_{t}
\dot{\omega}_{0}\left(\beta\frac{\varphi_{1}-x}{\varepsilon}\right)
\omega_{0}\left(\beta\frac{x-\varphi_{2}}{\varepsilon}\right)\\
&\qquad-\left.\left[\beta(x-\varphi_{2})\right]_{t}
\dot{\omega}_{0}\left(\beta\frac{x-\varphi_{2}}{\varepsilon}\right)
\omega_{0}\left(\beta\frac{\varphi_{1}-x}{\varepsilon}\right)\right\}\notag,
\end{align}
\begin{align}\label{UdifX:1}
&\check{u}_{x}=\frac{\beta}{2\varepsilon}\left\{
-\dot{\omega}_{0}
\left(\beta\frac{\varphi_{1}-x}{\varepsilon}\right)+
\dot{\omega}_{0}\left(\beta\frac{x-\varphi_{2}}{\varepsilon}\right)
\right.\\
&\qquad\left.+\dot{\omega}_{0}
\left(\beta\frac{\varphi_{1}-x}{\varepsilon}\right)
\omega_{0}\left(\beta\frac{x-\varphi_{2}}{\varepsilon}\right)-
\dot{\omega}_{0}\left(\beta\frac{x-\varphi_{2}}{\varepsilon}\right)
\omega_{0}\left(\beta\frac{\varphi_{1}-x}{\varepsilon}\right)\right\}
\notag.
\end{align}
Now we group the terms in the subintegral expression so that
one group contain terms with the factor
$\dot\omega_0(\beta(\varphi_i-x)/\varepsilon)$,
and the other group contain terms with the factor
$\dot\omega_0(\beta(x-\varphi_2)/\varepsilon)$.
Thus, we write the integral as the sum of two integrals.
Then we perform the change of variables
\begin{equation}
x\to\varphi_1-\frac{\varepsilon z}{\beta}
\end{equation}
in the first integral
and the change of variables
\begin{equation}
x\to\varphi_2+\frac{\varepsilon z}{\beta}
\end{equation}
in the second integral.

From (68) and (69), we obtain
\begin{align}
&\int u_t u_x\zeta(x)\,dx
\\
&\quad =-\frac12\bigg[\varphi_{1t}+\frac{\beta_\tau}{\beta^2}\psi'_0\bigg]
\zeta(\varphi_1)[(1+z)(\Omega'_{z}-\Omega'_\eta)
+(1-z-\eta)\Omega'_{\eta}][\Omega'_\eta-\Omega'_z]\,dz
\notag\\
&\quad
-\frac12\bigg[\varphi_{2t}+\frac{\beta_\tau}{\beta^2}\psi'_0\bigg]
\zeta(\varphi_2)[(1+z)\Omega'_{1z}
+(1-z-\eta)\Omega'_{\eta}][\Omega'_\eta-\Omega'_z]\,dz
\notag\\
&\quad
+\mathcal{O}(\varepsilon).
\notag
\end{align}

Here we used the formula for the derivative
$\beta_t=\varepsilon^{-1}\beta_\tau\psi'_{0t}$
and the fact that the functions in the integrand in~(72)
are invariant under the change $z\to-z-\eta$,
since each bracket is multiplied by~$-1$
in this change.

Now we calculate the other expressions contained in (6).
We have
\begin{align}
&\frac{\varepsilon}{2}\int \zeta'(x)(\check u_x)^2\,dx
=\frac{\beta^2}{8\varepsilon}\int \zeta'(x)
\bigg[-\dot{\omega}_{0}
\left(\beta\frac{\varphi_{1}-x}{\varepsilon}\right)
+\dot{\omega}_{0}\left(\beta\frac{x-\varphi_{2}}{\varepsilon}\right)
\\
&\qquad
+ \dot{\omega}_{0} \left(\beta\frac{\varphi_{1}-x}{\varepsilon}\right)
\omega_{0}\left(\beta\frac{x-\varphi_{2}}{\varepsilon}\right)
- \dot{\omega}_{0}\left(\beta\frac{x-\varphi_{2}}{\varepsilon}\right)
\omega_{0}\left(\beta\frac{\varphi_{1}-x}{\varepsilon}\right)\bigg]^2\,dx
\notag\\
&=\frac{\beta}{8}\big(\zeta'(\varphi_1)+\zeta'(\varphi_2)\big)
\int\big\{\dot\omega_0(z)(1-\omega_0(-z-\eta))^2
\notag\\
&\qquad
-\dot\omega_0(z)\dot\omega_0(-z-\eta)
(1-\omega_0(z))(1-\omega_0(-z-\eta)) \big\}\,dz
+\mathcal{O}(\varepsilon).
\notag
\end{align}
It is easy to see that the integrand in the right-hand side
of (73) can be written as
\begin{align*}
&\int\{(\dot\omega_0(z)(1-\omega_0(-z-\eta))^2
\\
&\qquad
-\dot\omega_0(z)\dot\omega(-z-\eta)
(1-\omega_0(z))(1-\omega_0(-z-\eta))\}\,dz
=\frac14\int(\Omega'_z)^2\,dz.
\end{align*}
Thus, we finally obtain
$$
\frac{\varepsilon}{2}\int\zeta'(\check{u}_x)^2\,dx
=\frac{\beta}{4}(\zeta'(\varphi_1)+\zeta'(\varphi_2))
\int(\Omega'_z)^2\,dz+\mathcal{O}(\varepsilon).
$$
Then the left-hand side of (73) is positive,
and hence the integrand expression in the right-hand side
is also positive (until the measure of the points at which
$|u_x|\geq\mbox{const}$ is positive).

The last term in (6) can be considered similarlly:
$$
\frac1{\varepsilon}\int F(\check u)\zeta'_x\,dx
=\frac{\beta^{-1}}{2}(\zeta'(\varphi_1)+\zeta'(\varphi_2))
\int F(\Omega)\,dz
+\mathcal{O}(\varepsilon).
$$

To prove this relation, we must successively perform
the change of variables (71) in the integral
in the left-hand side
and then consider the half-sum of the integrals obtained,
which, obviously, is equal to the original integral
in the left-hand side.
Moreover, we must take into account
the estimates (34), (35), and 
$$
\Omega(z,\eta)
=\begin{cases}
1+\mathcal{O}(\exp(-2z)), &z\to\infty,
\\
\omega_0(-z-\eta)+\mathcal{O}(\exp(2z)), &z\to-\infty,
\end{cases}
$$
which together with the explicit form of the function
$F(u)=\frac{u^4}{4}-\frac{u^2}{2}+\frac14$
imply that the integrals
$$
\int zF(\Omega)\,dz
$$
converge.
\end{proof}

\end{document}